\documentclass[a4paper,11pt]{article}
\pdfoutput=1 

\usepackage{jheppub} 
\usepackage[T1]{fontenc} 
\usepackage[numbers]{natbib}
\usepackage[italian,english]{babel}
\usepackage{hyperref}
\usepackage{ifpdf}
\usepackage{subfigure}
\usepackage{slashed}
\usepackage{braket}
\usepackage{pgffor}
\usepackage{amssymb}
\usepackage{amsfonts}
\usepackage{epsf}
\usepackage{rotating}
\usepackage{graphicx}
\usepackage{amsmath}
\usepackage{fancyhdr}
\usepackage{lineno}
\usepackage{babel}
\usepackage{graphics}
\usepackage{pstricks}
\usepackage{color}
\usepackage{multirow}
\usepackage{float}
\usepackage{lineno}
\usepackage{mathtools}
\usepackage{stackengine}
\usepackage{comment}

\newcommand{\lsim}{\mathrel{\mathop{\kern 0pt \rlap
{\raise.2ex\hbox{$<$}}}
\lower.9ex\hbox{\kern-.190em $\sim$}}}
\newcommand{\gsim}{\mathrel{\mathop{\kern 0pt \rlap
{\raise.2ex\hbox{$>$}}}
\lower.9ex\hbox{\kern-.190em $\sim$}}}

\newcommand{\be}{\begin{equation}}
\newcommand{\ee}{\end{equation}}
\newcommand{\bea}{\begin{eqnarray}}
\newcommand{\eea}{\end{eqnarray}}

\begin{document}
\title{Higgs Scattering and Entanglement in SMEFT}
\author[a]{Hyun Min Lee}
\author[a]{, Rojalin Padhan}
\affiliation[a]{Department of Physics, Chung-Ang University, Seoul 06974, Korea}
\emailAdd{hminlee@cau.ac.kr}
\emailAdd{rojalinpadhan2014@gmail.com}


\abstract{We regard the weak isospin of the Higgs doublet as a qubit and classify the entanglement measures for the Higgs scattering in the Standard Model Effective Field Theories (SMEFT) and their Ultra-Violet complete models. We consider Higgs scattering in the unbroken phase of electroweak symmetry. Treating the final state as a momentum-isospin bipartite system, we obtain von Neumann and linear entropies to quantify momentum-isospin correlation. From the momentum reduced state, we calculate the concurrence, which measures the entanglement between the two isospins. Both quantities are set by the isospin singlet and triplet scattering amplitudes, and hence by the  Wilson coefficients of the dimension-6 and dimension-8 Higgs operators. We find that the von Neumann entropy grows as a function of the total energy in SMEFT as compared to the SM case, but it undergoes a cancellation in the medium energy below the cut-off scale due to the interference effects between the dimension-4 and dimension-8 operators, in particular, when the effective interactions stem dominantly from a massive graviton. 
Assuming the dominance of dimension-8 operators, we find the conditions for entanglement suppression/generation in the forward or backward scatterings or across all the kinematics. We also show the correlations between the entanglement suppression and the positivity bounds in the forward limit. 
}

\maketitle              

\section{Introduction}
The interplay between quantum information theory and high energy physics has attracted considerable attention in recent years. In particular, quantum entanglement has emerged as a useful tool for probing scattering processes, since the quantum states of the outgoing particles can carry information about the underlying dynamics. The recent proposal~\cite{Afik:2020onf} and observation of quantum entanglement in top-quark pair production by the ATLAS and CMS collaborations~\cite{ATLAS:2023fsd,CMS:2024pts} have demonstrated that such observables are experimentally accessible at the LHC, thereby motivating studies of entanglement effects in other scattering processes.

While entanglement is most commonly discussed in the context of spin correlations, it can also arise among other quantum numbers such as flavor, isospin, and momentum. 
An interesting development in this direction is the observation that entanglement suppression in low energy scattering may be linked to the emergence of enhanced symmetries that are not manifest in the underlying Lagrangian.
This idea was first explored in nucleon-nucleon scattering~\cite{Beane:2018oxh}, where entanglement suppression in strong interaction was found to coincide with the emergence of the approximate spin-flavor symmetry. The framework was later revisited in several papers.~\cite{Low:2021ufv,Liu:2022grf, Hu:2025lua}.  Further, it has been established that minimally entangled scattering amplitudes are closely related to the Identity and SWAP operators, leading to symmetry structures such as Wigner's spin-flavor symmetry and non-relativistic conformal invariance in low-energy QCD~\cite{Low:2021ufv}. 

Considering flavor as a qubit,  quantum entanglement has also been explored through scalar scattering~\cite{Carena:2023vjc,Kowalska:2024kbs,Busoni:2025dns,Kowalska:2025qmf} where the authors investigated entanglement suppression in two Higgs doublet models (2HDMs), identifying regions of parameter space where scalar scattering exhibits either maximal or minimal entanglement. A general framework has been developed~\cite{McGinnis:2025brt}, showing that restricting the S-matrix to minimally entangling operators is equivalent to the emergence of an $SU(N)$ global symmetry. 
A similar study in $2\to2$ quark scattering~\cite{Thaler:2024anb} shows that minimum entanglement implies an almost diagonal Cabibbo-Kobayashi-Maskawa (CKM) matrix. In
ref.~\cite{Aoude:2024xpx}, the authors established a connection between positivity bounds on forward scattering amplitudes and the entanglement generated during the scattering process, providing a new way to interpret amplitude positivity in terms of quantum information.

While quantum entanglement has been studied through scalar scattering in the context of 2HDMs, a similar study in the Standard Model (SM) Higgs sector at the level of the Standard Model Effective Field Theory (SMEFT) is still lacking. The dimension-8 derivative operators in the Higgs sector are of particular 
interest, since (i) they generate non-trivial momentum dependence in the scattering amplitudes, (ii) they are constrained by positivity bounds from forward elastic scattering~\cite{Adams:2006sv,Bi:2019phv,Remmen:2019cyz,Zhang:2021eeo,Yamashita:2020gtt,Kim:2023pwf,Kim:2023bbs} and (iii) they are accessible at the LHC through 
processes such as longitudinal vector-boson scattering. The 
interplay between entanglement suppression conditions and positivity bounds at this 
order has not been worked out.

In this work, we investigate isospin entanglement in $2\to 2$ SM Higgs scattering 
within the SMEFT and some Ultra-Violet (UV) complete models, treating the weak isospin of each external Higgs leg 
as a qubit. We closely follow the entanglement formalism presented in ref. ~\cite{Kowalska:2024kbs,Kowalska:2025qmf}. We consider both the $HH \to HH$ and $H\tilde{H} \to H\tilde{H}$ 
channels with ${\tilde H}=i\sigma^2 H^*$ and compute the singlet and triplet amplitudes from dimension-8 operators in the Higgs basis. We identify the Wilson coefficient relations 
that suppress isospin entanglement in the forward and backward kinematic 
limits, and compare these to the positivity bounds. We find that entanglement suppression conditions correspond to specific boundaries of the region allowed by the positivity of the amplitude.

The paper is organised as follows.
In Section.~\ref{sec.EntMeasures}, we outline various entanglement measures in a bipartite system. 
In Section~\ref{sec:formalism}, we review the formalism to investigate the entanglement in scalar scattering.
In Section~\ref{sec:framework}, we set 
up the qubit framework for isospin entanglement and derive the general 
form of the $SU(2)_L$ invariant amplitude matrix.  Section~\ref{sec:Entropy} 
analyses the entanglement entropy as a measure of isospin-momentum entanglement,  followed by Section~\ref{sec:Concurence}, where we discuss concurrence to quantify entanglement between the two isospin qubits. Section~\ref{sec:positivity} 
compares the entanglement suppression conditions to the positivity bounds. 
We conclude in Section~\ref{sec:conclusions}.
There are two appendices listing the Feynman rules for the Higgs scattering and showing the derivation of positivity bounds in the complex basis for the Higgs doublet.

\section{Summary of entanglement measures in bipartite system}
\label{sec.EntMeasures}
\begin{itemize}
    \item {Density Matrix and Purity: } For a quantum state described by a density operator $\rho$ acting on a
Hilbert space $\mathcal{H}$, the density operator satisfies $\rho \geq 0$,
$\rho^\dagger = \rho$, and $\mathrm{Tr}(\rho) = 1$.
The purity of a state is measured by $\mathrm{Tr}(\rho^2)$,
with $\mathrm{Tr}(\rho^2) = 1$ for a pure state and
$\mathrm{Tr}(\rho^2) < 1$ for a mixed state.
For a $d$-dimensional system, the bounds are
$    \frac{1}{d} \leq \mathrm{Tr}(\rho^2) \leq 1$.

\item{von Neumann Entropy:}
The von Neumann entropy was introduced by von Neumann~\cite{vonNeumann1932}. It is defined
as~\cite{vonNeumann1932, Nielsen2000}
\begin{equation}
 \mathcal{S}_{\rm vN}(\rho) = -\mathrm{Tr}(\rho \log_2 \rho) = -\sum_i \lambda_i \log_2 \lambda_i,    
\end{equation}
where $\lambda_i$ are the eigenvalues of $\rho$. For a bipartite pure state
$|\psi\rangle_{AB}$, Bennett et al.~\cite{Bennett1996} showed that a natural
measure of the entanglement is the von Neumann entropy of either subsystem:
$    E = \mathcal{S}_{\rm vN}(\rho_A) = \mathcal{S}_{\rm vN}(\rho_B)$,
where $\rho_A = \mathrm{Tr}_B(|\psi\rangle \langle\psi|)$ is the reduced density
matrix of the subsystem $A$.
\item{Linear Entropy:}
The linear entropy is defined as~\cite{Zanardi2001, Pauletti2023}
\begin{equation}
\mathcal{S}_L(\rho) = 1 - \mathrm{Tr}(\rho^2)    
\end{equation}
which arises from linearizing the logarithm in the von Neumann entropy. It is also a special case of the Tsallis entropy of order 2~\cite{Tsallis1988}. For a
$d$-dimensional system:
$    0 \leq \mathcal{S}_L(\rho) \leq \frac{d-1}{d}$,
with $\mathcal{S}_L = 0$ for a pure state and $\mathcal{S}_L = (d-1)/d$ for the maximally mixed
state. The linear entropy is 
qualitatively equivalent to the von Neumann entropy and is used 
to quantify entanglement~\cite{Pauletti2023}.
As an entanglement measure for a pure bipartite state, the entanglement in terms of linear
entropy is defined as $
    E_L = \mathcal{S}_L(\rho_A) = 1 - \mathrm{Tr}(\rho_A^2)$.
\item{Concurrence:}
The concurrence was introduced by Hill and Wootters~\cite{Hill1997,Wootters1998}, as an entanglement measure for two-qubit
systems. For a pure two-qubit state $|\psi\rangle=\sum_{i,j} c_{ij}\, |i\rangle_A |j\rangle_B,\quad i,j\in\{1,2\}$, it is defined as
\begin{equation}
    C(|\psi\rangle) = \left|\langle\tilde{\psi}|\psi\rangle\right|=2\left|c_{11}c_{22}-c_{12}c_{21}\right|,
\end{equation}
where $|\tilde{\psi}\rangle = (\sigma_y \otimes \sigma_y)|\psi^*\rangle$ is the
spin-flipped state. For a mixed two-qubit state $\rho$, the concurrence is defined as 
\begin{equation}
    C(\rho) = \max\!\left\{0,\; \sqrt{\lambda_1} - \sqrt{\lambda_2} -
    \sqrt{\lambda_3} - \sqrt{\lambda_4}\right\},
\end{equation}
where $\lambda_i$ are the eigenvalues in decreasing order of the non-Hermitian
operator~\cite{Wootters1998}
\begin{equation}
    R = \rho\,(\sigma_y \otimes \sigma_y)\,\rho^*\,
    (\sigma_y \otimes \sigma_y).
\end{equation}
The concurrence  $C = 0$ for a separable state and $C = 1$ for a
maximally entangled state.
\item von Neumann entropy vs
linear entropy: 
For a qubit, we can relate von Neumann entropy and linear entropy. We can write the two eigenvalues of the reduced density matrix $\rho_A$ as
$\lambda_{\pm}=\frac{1\pm r}{2},\quad 0\le r\le 1.
$
The von Neumann entropy can be written as 
\bea
 \mathcal{S}_{\rm vN}(\rho_A)
=
-\frac{1+r}{2}\log_2\frac{1+r}{2}
-\frac{1-r}{2}\log_2\frac{1-r}{2}.
\eea

and linear entropy is given by
\bea
\mathcal{S}_{L}(\rho_A)
=
1-\mathrm{Tr}(\rho_A^2)=1-\frac{1+r^2}{2}=\frac{1-r^2}{2}.
\eea

Replacing $r=\sqrt{1-2\mathcal{S}_L(\rho_A)}$ in the von Neumann entropy gives
\bea
\mathcal{S}_{\rm vN}(\rho_A)
=
-\frac{1+\sqrt{1-2\mathcal{S}_L}}{2}
\log_2
\left(
\frac{1+\sqrt{1-2\mathcal{S}_L}}{2}
\right) \nonumber \\
\\-\frac{1-\sqrt{1-2\mathcal{S}_L}}{2}
\log_2
\left(
\frac{1-\sqrt{1-2\mathcal{S}_L}}{2}
\right).
\eea
For nearly pure states, $\mathcal{S}_L\ll1$, we define $r=1-\epsilon,\quad \epsilon\ll1$
Using
\[
\mathcal{S}_L(\rho_A)=\frac{1-r^2}{2} = \epsilon-\frac{\epsilon^2}{2} \approx \epsilon,\]
the entropy can be expanded as
\bea
\mathcal{S}_{\rm vN}(\rho_A)
\approx
\frac{\mathcal{S}_L(\rho_A)}{\ln2}
\left[
1-\ln\left(\frac{\mathcal{S}_L(\rho_A)}{2}\right)
\right].
\eea
\item Concurrence vs Linear entropy:
The concurrence for a pure two-qubit state is given by
\bea
C=2\sqrt{\det\rho_A}=2\sqrt{\lambda_+\lambda_-}=\sqrt{1-r^2}.
\eea
Using the expression for the linear entropy,
we obtain the relation
\bea
C=\sqrt{2 \mathcal{S}_L(\rho_A)}
\eea
\end{itemize}
\section{The formalism for entanglement in scattering } \label{sec:formalism}
In this section, we revisit the formalism~\cite{Kowalska:2024kbs} to study entanglement in a $2\to2$ scalar scattering process. We consider the scattering $H_\alpha H_\beta \to H_\gamma H_\delta$. Here, the quantum state of the scalar $H_\alpha$ is defined by momentum degrees of freedom and a discrete quantum number (flavor in general) as 
$\mid \mathbf{p}, \alpha\rangle =|\mathbf{p}\rangle \otimes |\alpha\rangle$, where $|p\rangle$ and $|\alpha\rangle$  denote momentum and flavor states, respectively. $|\mathbf{p}\rangle$ span the momentum Hilbert space $L^2(\mathbb{R}^3)$ and  $|\alpha\rangle$ span a 2-dimensional Hilbert space  $\mathbb{C}^2$. The quantum state $\mid \mathbf{p}, \alpha\rangle$ satisfies the normalization condition
\begin{equation}    
\label{eq:1pNorm}
\langle \mathbf{p}', \beta \mid \mathbf{p}, \alpha \rangle 
= (2\pi)^3 \, 2E_p \, \delta^{(3)}(\mathbf{p} - \mathbf{p}') \, \delta_{\alpha \beta},\qquad\alpha, \beta=1,2
\end{equation}
The full Hilbert space of the two scattering particles, which consists of the momentum Hilbert space, $\mathcal{H}_{\mathbf{p}}$ and  flavor Hilbert space $\mathcal{H}_f$  is defined as 
\begin{equation}
\mathcal{H}_{\textrm{tot}}=\mathcal{H}_{\mathbf{p}} \otimes \mathcal{H}_{f}= (L^2(\mathbb{R}^3) \otimes  L^2(\mathbb{R}^3)) \otimes ( \mathbb{C}^2\otimes \mathbb{C}^2)= L^2(\mathbb{R}^3\otimes \mathbb{R}^3)\otimes \mathbb{C}^4\,.
\end{equation}
The bipartite quantum state in full Hilbert space $\mathcal{H}_{\textrm{tot}}$ is defined as $| \mathbf{p}_1 \mathbf{p}_2\rangle |\alpha \beta\rangle$
 with normalization conditions 
\begin{equation}     
 \begin{aligned}     
\label{eq:2pNorm}
\langle \mathbf{p}_1 \mathbf{p}_2\mid \mathbf{p}_1^{\prime} \mathbf{p}_2^{\prime} \rangle 
&= (2\pi)^6 \, 4E_1E_2 \, \delta^{(3)}(\mathbf{p}_1 - \mathbf{p}_1^\prime) \,\delta^{(3)}(\mathbf{p}_2 - \mathbf{p}_2^\prime) \, \\
\langle \alpha \beta\mid \gamma \delta \rangle &=\delta_{\alpha \gamma}  \, \delta_{\beta \delta}.
 \end{aligned}    
 \end{equation}
 As the total Hilbert space $\mathcal{H}_{\mathrm{tot}}$ consists of multiple subspaces, one can  bi-partition it in the following ways.

\paragraph{Bi-partition I:  Momentum vs.\  flavor:}
\begin{equation}
    \mathcal{H}_{\mathrm{tot}} =
    \underbrace{\mathcal{H}_\mathbf{p}}_{\text{subsystem A}} \otimes
    \underbrace{\mathcal{H}_{f}}_{\text{subsystem B}}
\end{equation}
Here, the total Hilbert space is decomposed as
$\mathcal{H}_{\mathrm{tot}}=\mathcal{H}_{\mathbf p}\otimes\mathcal{H}_{f}$
where $\mathcal{H}_{\mathbf p}$ and $\mathcal{H}_{f}$ denote the momentum and flavor Hilbert spaces, respectively. Each of these spaces further factorizes into the corresponding single-particle subspaces such as $ \mathcal{H}_\mathbf{p}=\mathcal{H}_\mathbf{p}^{(1)} \otimes \mathcal{H}_\mathbf{p}^{(2)}$. This choice of bi-partition allows us to study quantum correlations between momentum and flavor degrees of freedom and enables the computation of the associated entanglement entropy through the reduced density matrix of either sector.
\paragraph{Bi-partition II:  Particle 1 vs Particle 2:}
\begin{equation}
    \mathcal{H}_{\mathrm{tot}} =
    \underbrace{\mathcal{H}^{(1)}}_{\text{subsystem A}} \otimes
    \underbrace{\mathcal{H}^{(2)}}_{\text{subsystem B}}
\end{equation}
Alternatively, we can decompose as $\mathcal{H}_{\mathrm{tot}}=\mathcal{H}^{(1)}\otimes\mathcal{H}^{(2)}$, where $\mathcal{H}^{(i)}$ consists of both momentum and flavor degrees of freedom of particle $i$ as $\mathcal{H}^{(i)}=\mathcal{H}_\mathbf{p}^{(i)} \otimes
    \mathcal{H}_{f}^{(i)}$
In this bi-partition we can measure the entanglement between the two scattered particles. However, it is difficult to identify whether the two particles are entangled in momentum Hilbert space or  flavor Hilbert  space. Therefore, we follow bi-partition I in this work.

Following the analysis~\cite{Kowalska:2024kbs} we consider the initial two-particle state to be a pair of
plane waves with a highly collimated momentum distribution peaked around the incoming
momenta $\mathbf{p}_A$ and $\mathbf{p}_B$, namely
\begin{equation}
  |\text{in}\rangle = \frac{1}{\sqrt{V}}\, \sum_{\alpha,\beta=1}^2 a_{\alpha \beta}|\mathbf{p}_A \mathbf{p}_B\rangle |\alpha \beta \rangle,
  \qquad \sum_{\alpha,\beta}|a_{\alpha\beta}|^2 = 1 ,
\end{equation}
where the momentum part is an element of the plane-wave basis normalized through the
formal volume factor $V=\langle \mathbf{p}_A \mathbf{p}_B|\mathbf{p}_A \mathbf{p}_B\rangle$. This state is separable between the momentum and flavor qubit Hilbert spaces,
although the two flavor qubits may themselves be entangled. For the clarity of operational meaning of the two labelled subsystems, we remark that the subsystems
correspond to the two resolved outgoing momentum modes.
 
 After the scattering, the full final-state density matrix is
\begin{equation}
    \rho_{\mathrm{out}} = |\rm{out}\rangle\langle\rm{out}|,~~where ~|\rm{out}\rangle=S|\rm{in}\rangle
    \label{eq:rhoOut}
\end{equation}
Note that $ \rho_{\mathrm{out}}$ is normalized order-by-order in perturbation theory as a consequence of the  Optical theorem i.e. $ Tr(\rho_{\mathrm{out}})=1$. Furthermore, $ Tr(\rho_{\mathrm{out}}^2)=1$, implies that $|\rm{out}\rangle$ state is a pure state.

Following bipartition-I, we trace out the momentum degrees of freedom to obtain the reduced density matrix for the flavor subspace $\mathcal{H}_f$. Its matrix elements in the basis $\{|\alpha\beta\rangle\langle\gamma\delta|\}$   are given by~\cite{Kowalska:2024kbs} 
\begin{equation}\label{eq:Rho}
  \begin{split}
  \tilde{\rho}_{\alpha\beta,\gamma\delta}
  =&\int\int \frac{d^3 p_i}{(2\pi)^3}\frac{1}{2 E_i}\frac{d^3 p_j}{(2\pi)^3}\frac{1}{2 E_j}\,\rho_{\alpha\beta,\gamma\delta}(\mathbf{p}_i,\mathbf{p}_j;\mathbf{p}_i,\mathbf{p}_j)\\
 =& a_{\alpha\beta}\,a^{\ast}_{\gamma\delta}
+\Delta \left[-i\,  a_{\alpha\beta} \sum_{\epsilon\rho} \mathcal{M}^{\ast}_{\epsilon\rho,\gamma\delta}(p_A,p_B\to p_A, p_B) \,a^{\ast}_{\epsilon\rho}\right.  \\
&+\left.i\,a^{\ast}_{\gamma\delta}  \sum_{\epsilon\rho} \mathcal{M}_{\epsilon\rho,\alpha\beta}(p_A,p_B\to p_A, p_B) \,a_{\epsilon\rho}\right]\\
&+\Delta \int \int \frac{d^3 p_i}{(2\pi)^3}\frac{1}{2 E_i}\frac{d^3 p_j}{(2\pi)^3}\frac{1}{2 E_j} 
 (2\pi)^4 \delta^4(p_A+p_B-p_i-p_j)  \\
&\quad \times \sum_{\epsilon\rho,\tau\sigma} 
\mathcal{M}_{\epsilon\rho,\alpha\beta}(p_A,p_B\to p_i,p_j)
\mathcal{M}^{\ast}_{\tau\sigma,\gamma\delta}(p_A,p_B \to p_i,p_j) \,a_{\epsilon\rho} \,a^{\ast}_{\tau\sigma}\,.
  \end{split}
  \end{equation} 
Here, $\Delta=\delta(0)/ \left(4E_AE_B(2\pi)^2\delta^3(0)\right)$ is an indeterminate normalization factor.   
For identical Higgs final states, in order to avoid a double counting in tracing out the Higgs momenta, we need to divide the reduced density matrix in Eq.~(\ref{eq:Rho}) by a factor $2$ for $\gamma=\delta$, like we do for computing the cross sections. Moreover, the Bose symmetry is manifest from the symmetry of the reduced scattering amplitude under the exchange of two Higgs final states, namely, ${\cal M}_{\gamma\delta,\alpha\beta}(p_A, p_B\to p_i, p_j)={\cal M}_{\delta\gamma,\alpha\beta}(p_A, p_B\to p_j, p_i)$, as shown explicitly from the vertex Feynman rules for Higgs scattering in SMEFT in Appendix~\ref{feynrule}.

We note that positive definiteness of $\tilde{\rho}$ sets the limit $\Delta\le\Delta_{\rm max}$. For $\Delta<\Delta_{\rm max}$,  $Tr(\tilde{\rho}^2)<1$ and hence $\tilde{\rho}$ represents a mixed state in flavor space . While for  $\Delta=\Delta_{\rm max}$,  $\tilde{\rho}$  represents a pure state in flavor space. Further,  $Tr(\tilde{\rho}^2)<1$ indicates non-zero Linear entropy, which means momentum and flavor space are entangled.
For a generic scattering amplitude with angular dependence,
\begin{equation}
M(p_A,p_B;\theta)=g^2\,\hat M(\theta),
\end{equation}
where $g$ denotes the overall coupling strength and $\hat M(\theta)$ contains the normalized angular structure of the amplitude, $\Delta_{\rm max }$ is defined as~\cite{Kowalska:2025qmf}
\begin{equation}
\Delta_{\max} g^4
=
\frac{1}{16\pi}
\int d(\cos\theta)\,
|M(p_A,p_B;\theta)|^2 .
\end{equation}

Thus, $\Delta = \Delta_{\max}$ depends on the angular
distribution of the scattering amplitude and hence its value is theory dependent.  For an isotropic amplitude 
$\Delta_{\max}=\frac{1}{16\pi}$.
$\Delta = \Delta_{\max}$ corresponds to the scattered wave packet $\tilde\phi$ being parallel to the initial $\phi$. 

\section{Standard Model EFT and Higgs scattering amplitudes} \label{sec:framework}
 We regard the isospin quantum number of the SM Higgs doublet $H=[\,H^+,H^0\,]^T$ as qubit, with basis state $\{H_1, H_2\}=\{H^+,H^0\}=\{|1\rangle,|2\rangle \}$, where isospin eigenstates are assigned as $\{|I_3=+\frac{1}{2} \rangle,|I_3=-\frac{1}{2}\rangle\}=\{|1\rangle,|2\rangle \}$.

 \subsection{Higgs sector in the SMEFT and some UV completions}
 
In the limit of the unbroken electroweak symmetry, the effective Lagrangian relevant for the $H_\alpha H_\beta \to H_\gamma H_\delta$ scattering is as follows,
\begin{equation}\label{eq:scapot0}
\begin{aligned}
\mathcal{L}= \mathcal{O}_{4}+\mathcal{O}_{6}+\mathcal{O}_{8},
\,
\end{aligned}
\end{equation}
with
\begin{equation}
\begin{aligned}
\mathcal{O}_{4}=& ~-\mathcal{C}_{H4}\,(H^\dagger H)^2, \\
\mathcal{O}_{6}=& ~\dfrac{\mathcal{C}_{H4D2}}{\Lambda^2}\,(H^\dagger D^\mu H)^\star (H^\dagger D_\mu H) + \dfrac{\mathcal{C}_{H \Box}}{\Lambda^2}\,(H^\dagger H) \square (H^\dagger H),\\
\mathcal{O}_{8}=& ~\dfrac{\mathcal{C}_1}{\Lambda^4}\, (D_\mu H^\dagger D_\nu H) (D^\nu H^\dagger D^\mu H) + \dfrac{\mathcal{C}_2}{\Lambda^4}\, (D_\mu H^\dagger D_\nu H) (D^\mu H^\dagger D^\nu H)\\
+&\dfrac{\mathcal{C}_3}{\Lambda^4}\, (D_\mu H^\dagger D^\mu H) (D_\nu H^\dagger D^\nu H).
\end{aligned}
\end{equation}
Here, $C_{H_4}=\lambda_H$, namely, the Higgs quartic coupling in the SM. The Feynman rules for the 4-point Higgs vertices are given in the Appendix~\ref{feynrule}.
Following the LHC limit on the EFT operators in ref.~\cite{deBlas:2025hbr}, we can impose limits on $\mathcal{C}_{H\Box}$ by Higgs data, while $\mathcal{C}_{H4D2}$ is mainly constrained by EW precision observables, but not by Higgs data. Including the current Higgs data~\cite{deBlas:2025hbr} $95\%$ Highest Posterior Density Interval (HPDI) limit from the individual fit for $\mathcal{C}_{H\Box}$  is $\mathcal{C}_{H\Box}/\Lambda^2 \in \left[-0.49,0.44\right]~\rm {TeV}^{-2} $. Similar limit on $\mathcal{C}_{H4D2}$ from EW  data~\cite{deBlas:2025xhe}   is $\mathcal{C}_{H4D2}/\Lambda^2 \in \left[-0.02,0.02\right]~\rm {TeV}^{-2}$ (at $95\%$ HPDI limit from individual fit).\footnote{Note that this limit is approximate which is extracted from fig.10 of ref.~\cite{deBlas:2025xhe}.} Vector boson scattering can be used to constrain the dimension-8 operators \cite{CMS:2014mra,Zhang:2018shp}. The observed $95\%$ CL limit on the individual operators $\mathcal{C}_2$ and $\mathcal{C}_3$ are $\mathcal{C}_{2}/\Lambda^4 \in \left[-38,40\right]~\rm {TeV}^{-4}$ and $\mathcal{C}_{3}/\Lambda^4 \in \left[-118,120\right]~\rm {TeV}^{-4}$, respectively.

We refer to UV complete models leading to a particular set of the effective operators in the SMEFT:
\begin{itemize}
    \item Singlet scalar/Radion ($S$) \cite{Ellis:2023zim}: 
    \be
    \mathcal{C}_3=\frac{2\kappa^2_S}{\Lambda^2},\quad \mathcal{C}_1=\mathcal{C}_2= 0, \quad\mathcal{C}_{H \Box}=-\frac{1}{4} \,\mathcal{C}_3,\quad  \mathcal{C}_{H4D2}=0
    \label{eq:UV_singlet}
    \ee
    \item Triplet scalar ($\Sigma$) \cite{Corbett:2021eux,Ellis:2023zim}:  
    \be
    \mathcal{C}_1=-2 \,\mathcal{C}_3=\frac{4\kappa^2_\Sigma}{\Lambda^2}, \quad \mathcal{C}_2= 0, \quad \mathcal{C}_{H \Box}=-\frac{1}{4}\,\mathcal{C}_3, \quad \mathcal{C}_{H4D2}=-\frac{1}{2}\,\mathcal{C}_1
    \label{eq:UV_triplet}
    \ee
    \item Massive graviton ($G$) \cite{Kim:2023pwf,Kim:2023bbs}: 
    \be
    \mathcal{C}_1=\mathcal{C}_2=-\frac{3}{2} ~\mathcal{C}_3=c^2_H, \quad \mathcal{C}_{H \Box}=\frac{m^2_H}{3\Lambda^2}\, \mathcal{C}_1, \quad \mathcal{C}_{H4D2}=0
    \label{eq:UV_graviton}
    \ee
\end{itemize}
Here, $\kappa_S, \kappa_\Sigma, c_H$ denote the linear couplings of the new particles to the Higgs bilinear, the cutoff scales for singlet/radion, triplet and massive graviton are given by $\Lambda=M_S, M_\Sigma$ and $\sqrt{M_G M_*}$ (with $M_*/c_H$ being the inverse of the graviton coupling), respectively, and we ignored the $m^2_H/\Lambda^2$ corrections to the effective operators at the tree-level matching.
\subsection{The qubit basis with the SM Higgs}
The two-particle isospin space decomposes as $\mathbf{2} \otimes \mathbf{2} = \mathbf{1} \oplus \mathbf{3},$
into an isospin singlet ($I=0$) and triplet ($I=1$), so we can write the scattering amplitude for two isospin qubits as,
\begin{equation}
\mathcal{M} = \mathcal{M}_0 \, P_0 + \mathcal{M}_1 \, P_1,
\label{eq:projOP}
\end{equation}
where $P_0=(\mathbb{I}-\sigma\cdot\sigma)/4$ and $P_1=(3\mathbb{I}+\sigma\cdot\sigma)/4$ are  the singlet and triplet projectors, respectively. Projection operators in Eq.~(\ref{eq:projOP}) are also applicable to $S=1+i \mathcal{M}$.
From $P_0+P_1=1$, we obtain $S=S_0 P_0 + S_1 P_1$ with $S_{0,1}=1+i\mathcal{M}_{0,1}$.
Transforming to the computational basis $\{|11\rangle,|12\rangle, |21\rangle, |22\rangle\}$, we can write  the $SU(2)_L$ invariant scattering amplitude as
\begin{equation}
\mathcal{M} =
\begin{pmatrix}
\mathcal{M}_1 & 0 & 0 & 0 \\[2pt]
0 & \dfrac{\mathcal{M}_1 + \mathcal{M}_0}{2} & \dfrac{\mathcal{M}_1 - \mathcal{M}_0}{2} & 0 \\[6pt]
0 & \dfrac{\mathcal{M}_1 - \mathcal{M}_0}{2} & \dfrac{\mathcal{M}_1 + \mathcal{M}_0}{2} & 0 \\[6pt]
0 & 0 & 0 & \mathcal{M}_1
\end{pmatrix}.
\label{eq:Mmatrix}
\end{equation}

 To study entanglement properties of the scattering of two complex fields, $H_\alpha H_\beta \to H_\gamma H_\delta$, we consider the amplitude matrix $\mathcal{M}_{\alpha\beta,\gamma\delta}$
\begin{equation}
\label{eq:scat01}
\mathcal{M}({H_\alpha H_\beta \to H_\gamma H_\delta})=
\left(\begin{array}{cccc} 
m_{11} &0 &0 &0 \\ 
0 &m_{22} & m_{23}& 0 \\
0 & m_{32} & m_{33} & 0 \\
0 & 0 &0& m_{44}
\end{array}\right)
\end{equation}
where
\begin{equation}
\label{eq:ampRelation}
\begin{aligned}
    m_{11}=&\mathcal{M}_{11,11}=\mathcal{M}(H^+ H^+ \to H^+ H^+),\\
    m_{22}=&\mathcal{M}_{12,12}=\mathcal{M}(H^+ H^0 \to H^+ H^0),\\
    m_{23}=&\mathcal{M}_{12,21}=\mathcal{M}(H^+ H^0 \to H^0 H^+), \\
    m_{32}=&\mathcal{M}_{21,12}= m_{23}^*,\\
    m_{33}=&\mathcal{M}_{21,21}=\mathcal{M}(H^0 H^+ \to H^0 H^+)=m_{22}, \\
 m_{44}=&\mathcal{M}_{22,22}=\mathcal{M}(H^0 H^0 \to H^0 H^0) =m_{11}.
    \end{aligned}
\end{equation}
Here, the amplitudes for Higgs scattering in the high energy limit $s\gg m^2$ are given by
\bea
    {\cal M}_{11,11}&=
    &-4 \mathcal{C}_{H4}-
\frac{\mathcal{C}_{H4D2} \left(t+u\right)}{\Lambda ^2}
-\frac{2\mathcal{C}_{H\square} \left(t+u\right)}{\Lambda ^2}\nonumber  \\
&& + \frac{\mathcal{C}_1 \left(  t^2 + u^2\right) }{2\Lambda ^4}+\frac{\mathcal{C}_2 s^2}{\Lambda ^4}+\frac{ 
\mathcal{C}_3 \left(t^2 +u^2 \right)}{2\Lambda ^4},\\
 {\cal M}_{12,12}&=&   -2\mathcal{C}_{H4}-\frac{ \mathcal{C}_{H4D2} ~u}{\Lambda ^2}-\frac{2\mathcal{C}_{H\square}~ t}{\Lambda ^2}   +\frac{\mathcal{C}_1 ~u^2}{2 \Lambda ^4} +\frac{\mathcal{C}_2 ~s^2}{2 \Lambda ^4} 
        +\frac{\mathcal{C}_3 ~t^2}{2 \Lambda ^4}, \\    
     {\cal M}_{12,21}&=&  -2\mathcal{C}_{H4}-\frac{ \mathcal{C}_{H4D2} ~t}{\Lambda ^2}-\frac{2\mathcal{C}_{H\square}~ u}{\Lambda ^2}  +\frac{\mathcal{C}_1 ~t^2}{2 \Lambda ^4} +\frac{\mathcal{C}_2 ~s^2}{2 \Lambda ^4} 
        +\frac{\mathcal{C}_3 ~u^2}{2 \Lambda ^4} .
\eea
Note that the amplitudes here satisfy the relation $m_{11}=m_{22}+m_{23}$ similar to Eq.~(\ref{eq:Mmatrix}).
\subsection{The qubit basis with the SM Higgs and its complex conjugate}
\label{sec:AlterBasis}

We also consider an alternative basis for two qubits in the SM Higgs and its complex conjugate, namely, $\{H_\alpha,\tilde{H}_\beta \}$, with ${\tilde H}=i\sigma^2 H^*$. Thus, we list the components for the conjugate state by  $\{\tilde{H}_1,\tilde{H}_2\}=\{H^{0*},-H^-\}=\{H_2^{*},-H_1^*\}=\{|1\rangle,|2\rangle \}$.  In this case, the amplitude matrix ${\cal \hat M}_{\alpha{\beta},\gamma{\delta}}$ for the scattering, $H_\alpha \tilde{H}_{\beta}\to H_\gamma \tilde{H}_{\delta}$,  is given by 
\bea
{\cal \hat M}(H_\alpha \tilde{H}_{\beta}\to H_\gamma \tilde{H}_{\delta})=
\left(\begin{array}{cccc} 
{\hat m}_{11} &0 &0 &0 \\ 
0 &{\hat m}_{22} &  {\hat m}_{23}& 0 \\
0 & {\hat m}_{32} & {\hat m}_{33} & 0 \\
0 & 0 &0& {\hat m}_{44}
\end{array}\right)
\label{eq:AmpHHc}
\eea
where ${\hat m}_{ij}$  is related to the $\mathcal{M}_{\alpha\beta,\gamma\delta}$ by crossing symmetry as
\bea
{\hat m}_{11}&=& {\cal \hat M}_{1 1,1 1}={\cal \hat M}(H^+ H^{0*}\to H^+ H^{*0})={\cal M}_{12,12}(s\leftrightarrow u,t), \nonumber \\
{\hat m}_{22}&=&{\cal \hat M}_{12,12}={\cal \hat M}(H^+ H^-\to H^+ H^-)={\cal M}_{11,11}(s\leftrightarrow u,t),  \nonumber \\
{\hat m}_{23}&=&{\cal \hat M}_{12,21}={\cal \hat M}(H^+ H^-\to H^0 H^{0*})=-{\cal M}_{12,21}(s\leftrightarrow u,t),  \nonumber \\
{\hat m}_{32}&=&{\cal \hat M}_{21,12}={\hat m}_{23}^*,  \nonumber \\
{\hat m}_{33}&=& {\cal \hat M}_{21,21}={\cal \hat M}(H^0 H^{0*}\to H^0 H^{0*})={\cal M}_{22,22}(s\leftrightarrow u,t), \nonumber  \\
{\hat m}_{44}&=& {\cal \hat M}_{22,22}={\cal \hat M}(H^0 H^-\to H^0 H^-)={\cal M}_{21,21}(s\leftrightarrow u,t).
\eea
In the above equation, the minus sign in the amplitude ${\hat m}_{23}$ arises from the field definition $\{\tilde{H}_1,\tilde{H}_2\}=\{H^{0*},-H^-\}$.  Here,  the amplitudes for Higgs scattering in the limit $s\gg m^2$ are given by
\bea
 {\cal \hat M}_{11,11}&=&  -2\mathcal{C}_{H4}-\frac{ \mathcal{C}_{H4D2} ~s}{\Lambda ^2}-\frac{2\mathcal{C}_{H\square}~ t}{\Lambda ^2} +\frac{\mathcal{C}_1 ~s^2}{2 \Lambda ^4} +\frac{\mathcal{C}_2 ~u^2}{2 \Lambda ^4} 
        +\frac{\mathcal{C}_3 ~t^2}{2 \Lambda ^4} \\
 {\cal \hat M}_{12,12}&=
    &-4 \mathcal{C}_{H4}-
\frac{\mathcal{C}_{H4D2} \left(s+t\right)}{\Lambda ^2}
-\frac{2\mathcal{C}_{H\square} \left(s+t\right)}{\Lambda ^2}\nonumber  \\
  &&+ \frac{\mathcal{C}_1 \left(  s^2 + t^2\right) }{2\Lambda ^4}+\frac{\mathcal{C}_2 u^2}{\Lambda ^4}+\frac{ 
\mathcal{C}_3 \left(s^2 +t^2 \right)}{2\Lambda ^4}\\
 {\cal \hat M}_{12,21}&=&  2\mathcal{C}_{H4}+\frac{ \mathcal{C}_{H4D2} ~t}{\Lambda ^2}+\frac{2\mathcal{C}_{H\square}~ s}{\Lambda ^2}   -\frac{\mathcal{C}_1 ~t^2}{2 \Lambda ^4} -\frac{\mathcal{C}_2 ~u^2}{2 \Lambda ^4} 
        -\frac{\mathcal{C}_3 ~s^2}{2 \Lambda ^4}
    .
\eea
\section{Higgs scattering and entanglement entropy} \label{sec:Entropy}
In this section, we discuss the entanglement generated in the scattering process, as encoded in the final state density matrix 
$\rho_{\rm out}$ in Eq.~(\ref{eq:rhoOut}). As mentioned before, $\rho_{\rm out}$ indicates a pure state; therefore, we can quantify the entanglement between the momentum and isospin Hilbert space using the von Neumann Entropy or the Linear Entropy. We refer to entanglement between the isospin and momentum degrees of freedom as isospin-momentum entanglement.

 Before going into the detailed calculation, we can comment on the impact of scattering based on whether the initial state is an isospin singlet/triplet state or an admixture of both. The computational basis states $|12\rangle$ and $|21\rangle$  are  admixture of  isospin singlet and  isospin triplet state as $|12\rangle=1/\sqrt{2}\left(|I=0\rangle +|I=1,I_3=0\rangle\right) $ and $|21\rangle=1/\sqrt{2}\left(-|I=0\rangle +|I=1,I_3=0\rangle \right)$ whereas $|11\rangle=|I=1,I_3=1\rangle $ and $|22\rangle=|I=1,I_3=-1\rangle$ are isospin triplet states. 
A generic two-particle isospin state $|\Psi\rangle = \sum_{\alpha,\beta=1}^2 a_{\alpha\beta}|\alpha\beta\rangle$
can be written in terms of isospin eigenstates as
\begin{align}|\Psi\rangle &=\tfrac{a_{12} - a_{21}}{\sqrt{2}}|I=0\rangle  \nonumber\\
&+ a_{11} |I=1,I_3=1\rangle + \tfrac{a_{12} + a_{21}}{\sqrt{2}}|I=1,I_3=0\rangle 
+ a_{22} |I=1,I_3=-1\rangle .
\end{align}

 Note that depending on the value of $a_{\alpha\beta}$, the state  $|\Psi\rangle$ can be an isospin triplet/singlet state or a mixture of triplet and singlet states. It is an isospin triplet state if $ a_{12} = a_{21}$. For an initial state
 \begin{equation}
  |\text{in}\rangle = \frac{1}{\sqrt{V}}\, \sum_{\alpha,\beta=1}^2 a_{\alpha \beta}|\mathbf{p}_A \mathbf{p}_B\rangle \otimes |\alpha \beta \rangle,
  \qquad \sum_{\alpha,\beta}|a_{\alpha\beta}|^2 = 1 ,
\end{equation}
which is separable across the isospin-momentum bipartition, we can write the final state as  
\begin{align}
|{\rm out}\rangle=&(I+i\mathcal{M})  |\rm in\rangle \nonumber \\
=&|{\rm in}\rangle + \frac{i}{\sqrt{V}}\frac{a_{12} - a_{21}}{\sqrt{2}}|I=0\rangle  \otimes \mathcal{M}_0|\mathbf{
p}_A \mathbf{p}_B\rangle \nonumber \\
+ &\frac{i}{\sqrt{V}}\left( a_{11} |I=1,I_3=1\rangle + \frac{a_{12} + a_{21}}{\sqrt{2}}|I=1,I_3=0\rangle 
+ a_{22} |I=1,I_3=-1\rangle\right) \otimes \mathcal{M}_1|\mathbf{
p}_A \mathbf{p}_B\rangle 
\label{Eq:CtoI}
\end{align}
As the isospin singlet and triplet subspaces are individually preserved, initial entanglement of any initial state which is  isospin singlet state  or isospin triplet state  remains unchanged after scattering. If the initial state $|\rm in\rangle$ is an isospin singlet/triplet  state, the outgoing state remains separable across the isospin-momentum bipartition. Consequently, the von Neumann entropy, which quantifies isospin-momentum entanglement, is zero after scattering. Non-zero isospin-momentum entanglement is generated when  $|\rm in\rangle$ is an admixture of both isospin singlet and triplet states. Entanglement suppression is obtained if $\mathcal{M}_0 = \mathcal{M}_1$ across all the kinematics.

In the following, we calculate the reduced density matrix $\tilde{\rho}$  and isospin-momentum entanglement for various initial two-particle states.
\subsection{Higgs basis: Product state 1}
\label{sec:higgsPS1}
Here we consider an initial isospin state $|12\rangle =1/\sqrt{2} ~(|I=0\rangle +|I=1,I_3=0\rangle )$ that is an equal admixture of the isospin singlet and triplet states. Therefore, we can  generate  non-trivial isospin-momentum entanglement  after scattering. In the computational basis we write this initial state  with  $a_{11}=0,a_{12}=1,a_{21}=0,a_{22}=0$, as,
\begin{equation}
|\text{in}\rangle = \frac{1}{\sqrt{V}}\,|\mathbf{p}_A \mathbf{p}_B\rangle |12\rangle.
 \end{equation}

 The reduced density matrix for this state derived from Eq.~(\ref{eq:Rho}) has the form
{\small
\begin{equation} \tilde{\rho}=
\begin{bmatrix}
0&0&0&0\\
 0&1+\Delta  \int d \Pi_2 M_{12,12}   M_{12,12}^*+\Delta  \left(i M_{12,12}^{\rm fw}-i M_{12,12}^{*\rm fw}\right) & \Delta  \int d \Pi_2 M_{12,12}   M_{12,21}^* -i \Delta M_{12,21}^{*\rm fw} & 0 \\
 0&\Delta  \int d \Pi_2 M_{12,12}^*   M_{12,21} +i \Delta M_{12,21}^{\rm fw} & \Delta  \int d \Pi_2 M_{12,21}   M_{12,21}^*&0\\
 0&0&0&0\\
\end{bmatrix}. 
\end{equation}
}
In the above $\mathcal{M}^{\rm fw}$ denotes the forward amplitude $\mathcal{M}(p_A p_B \to p_A p_B)$. Using Eq.~(\ref{eq:ampRelation}), we get 
\begin{equation} \tilde{\rho}=
\begin{bmatrix}
0&0&0&0\\
 0&1+\Delta  \int d \Pi_2 m_{22}   m_{22}^*+\Delta  \left(i m_{22}^{\rm fw}-i m_{22}^{*\rm fw}\right) & \Delta  \int d \Pi_2 m_{22}   m_{23}^* -i \Delta m_{23}^{*\rm fw} & 0 \\
 0&\Delta  \int d \Pi_2 m_{22}^*  m_{23} +i \Delta m_{23}^{\rm fw} & \Delta  \int d \Pi_2 m_{23}  m_{23}^*&0\\
 0&0&0&0\\
\end{bmatrix}. 
\end{equation}
Since $Tr(\tilde{\rho})=1+\Delta  \int d \Pi_2 m_{22}   m_{22}^*+\Delta  \int d \Pi_2 m_{23}   m_{23}^*$, the density matrix $\tilde{\rho}$ is not properly normalized. We therefore rescale it as  $\tilde{\rho}\to \tilde{\rho}/Tr(\tilde{\rho})$. The normalized $\tilde{\rho}$ up to $\mathcal{O}(\Delta)$ is given by
\begin{equation} \tilde{\rho}=
\begin{bmatrix}
0&0&0&0\\
 0&1-\Delta  \int d \Pi_2 |m_{23}|^2 & \Delta  \int d \Pi_2 m_{22} m_{23}^* -i \Delta m_{23}^{*\rm fw} & 0 \\
 0&\Delta  \int d \Pi_2 m_{22}^* m_{23}  +i \Delta m_{23}^{\rm fw} & \Delta  \int d \Pi_2 |m_{23}|^2 &0\\
 0&0&0&0\\
\end{bmatrix}.
\end{equation}
The two non-zero eigenvalues of $\tilde{\rho}$ are $\mathcal{E}$ and $1-\mathcal{E}$,   where
{\small
\begin{equation}
   \mathcal{E}=\frac{1}{2} + \frac{1}{2} \sqrt{1-4 \Delta \int d \Pi_2  |m_{23}|^2+4 \Delta ^2 |m_{23}^{\rm fw}|^2+4\Delta ^2  \left(\int d \Pi_2  |m_{23}|^2\right)^2+4\Delta ^2  \left(\int d \Pi_2  m_{22}m_{23}^*\right)^2}
   \label{eq:eigenV}
\end{equation}
}
The linear entropy is given by $\mathcal{S}_L=1-Tr(\tilde{\rho}^2)$ where  $Tr(\tilde{\rho}^2)=\mathcal{E}^2+(1-\mathcal{E})^2$.
 The von Neumann entropy, which quantifies the entanglement between momentum and isospin space, is defined as 
\begin{equation}
    \mathcal{S}_{\rm vN}= -\mathcal{E} \log_2(\mathcal{E})-(1-\mathcal{E}) \log_2(1-\mathcal{E}).
    \label{eq:vNentropy}
\end{equation}
The  two-body phase-space integrals  ($\int d\Pi_2\,|m_{23}|^2$ and $\int d\Pi_2\,m_{22} m_{23}^*$) including all operators are
\begin{equation}
\begin{aligned}
(16 \pi)\int d\Pi_2\,|m_{23}|^2&=
8 \mathcal{C}_{H4}^2-\frac{4s}{\Lambda ^2} \mathcal{C}_{H4}  (\mathcal{C}_{H4D2}+2 \mathcal{C}_{H \Box}) \\&
   +\frac{2 s^2}{3 \Lambda ^4} \left(\mathcal{C}_{H4D2}^2+2 \mathcal{C}_{H4D2} \mathcal{C}_{H \Box}-2 \mathcal{C}_{H4} (\mathcal{C}_1+3 \mathcal{C}_2+\mathcal{C}_3)+4 \mathcal{C}_{H \Box}^2\right)
   \\&+
   \frac{s^3}{6 \Lambda ^6} \left(\mathcal{C}_{H4D2} (3 \mathcal{C}_1+6 \mathcal{C}_2+\mathcal{C}_3)+2 \mathcal{C}_{H \Box} (\mathcal{C}_1+6 \mathcal{C}_2+3 \mathcal{C}_3)\right)\\&+
   \frac{s^4}{30 \Lambda ^8} \left(3 \mathcal{C}_1^2+3 \mathcal{C}_3^2+15 \mathcal{C}_2^2+\mathcal{C}_1 (10 \mathcal{C}_2+\mathcal{C}_3)+10 \mathcal{C}_2 \mathcal{C}_3\right),
   \end{aligned}
\end{equation}
\begin{equation}
\int d\Pi_2\,m_{22} m_{23}^*= \int d\Pi_2\,|m_{23}|^2-
    \frac{s^2}{192 \pi  \Lambda ^8} \left(2 \Lambda ^2 (\mathcal{C}_{H4D2}-2 \mathcal{C}_{H\Box})+ s ~(\mathcal{C}_1 -\mathcal{C}_3)\right)^2.
\end{equation}
\begin{figure}
    \centering
    \includegraphics[width=0.56\linewidth]{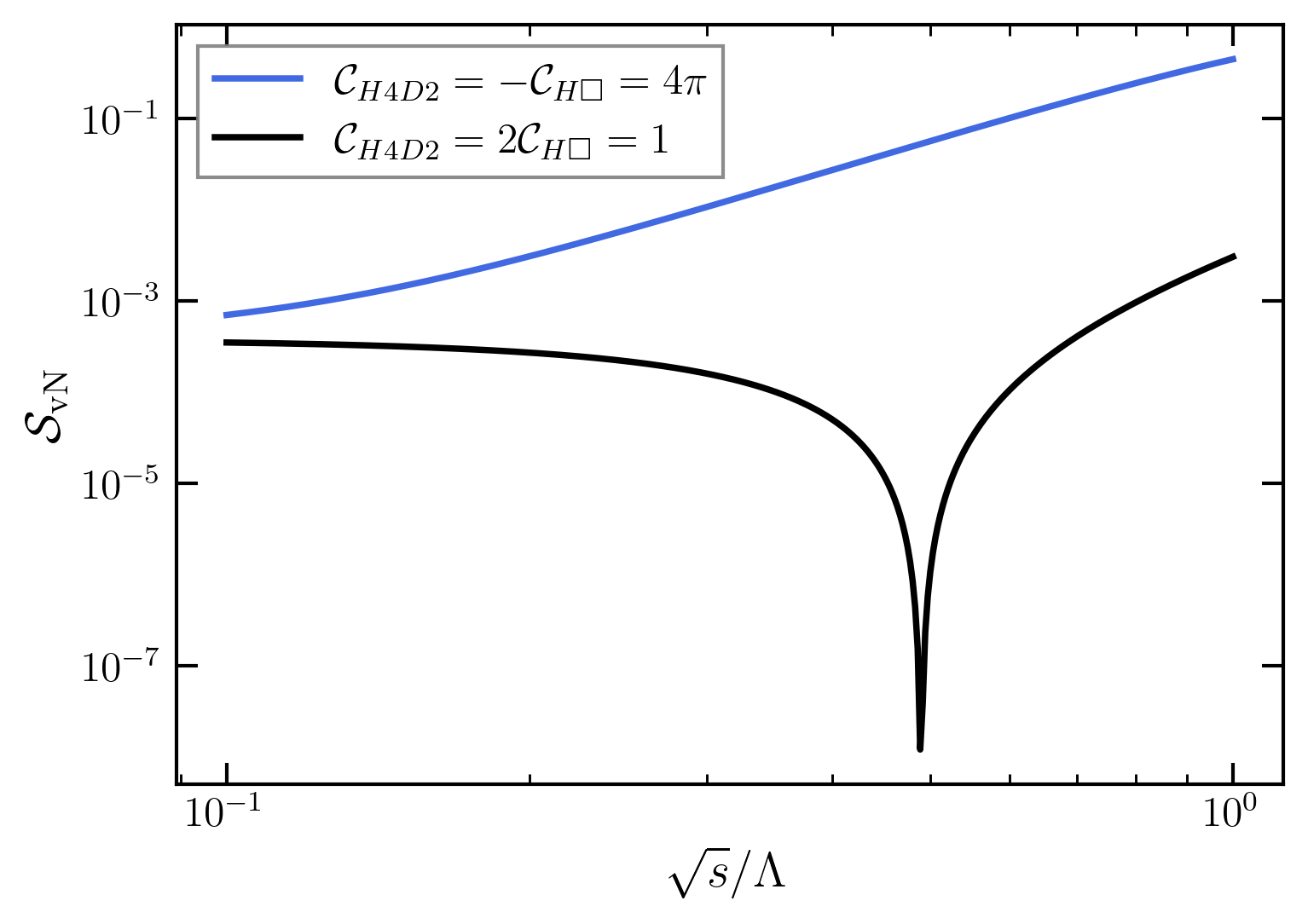}
    \caption{Variation of von Neumann entropy ($\mathcal{S}_{\rm vN}$) as a function of  $\sqrt{s}/\Lambda$ for two different combinations of $\mathcal{C}_{H\square}$ and $\mathcal{C}_{H4D2}$. Here we  fix $\Delta=0.9/16\pi$, $\mathcal{C}_{H4}=0.12$  and turn off dimension-8 operator coefficients, $\mathcal{C}_1=\mathcal{C}_2=  \mathcal{C}_3=0$.  }
    \label{fig:Entropy_d6}
\end{figure}

We show variation of the von Neumann entropy ($\mathcal{S}_{\rm vN}$) as a function of $\sqrt{s}/\Lambda$ in Fig.~\ref{fig:Entropy_d6}. Here we show a  marginal scenario, where the dimension-8 coefficients are switched off   $\mathcal{C}_1=\mathcal{C}_2=\mathcal{C}_3=0$ and $\Delta=0.9/(16\pi)$.
At leading order in $\Delta$ the entropy is a monotonic function of $\int d\Pi_2\,|m_{23}|^2$.
 If the  $\mathcal{M}_{\text{d-4}}\mathcal{M}_{\text{d-6}}$ interference term is negative\footnote{We denote amplitude from dimension-i operator as $\mathcal{M}_{\text{d-i}}$.}, three contributions in $\int d\Pi_2\,|m_{23}|^2$ add constructively and entropy  grows towards the maximum value $\mathcal{S}_{\rm vN}=1$. This feature is shown by the blue line where we consider $\mathcal{C}_{H4D2}=-\mathcal{C}_{H\Box}=4\pi$. The black line is for a specific condition $\mathcal{C}_{H4D2}=2\mathcal{C}_{H\Box}$, for which the singlet amplitudes $\mathcal{M}_0=m_{22}-m_{23}$ vanish and the momentum-isospin entanglement is controlled only by the triplet amplitude. At $\mathcal{O}(\Delta^2)$, the  entropy  depends  on  $\int d\Pi_2\,|m_{23}|^2 \propto (\mathcal{C}_{H\Box}\,s/\Lambda^2-\mathcal{C}_{H4})^2$.
At $\sqrt{s}/\Lambda=\sqrt{\mathcal{C}_{H4}/\mathcal{C}_{H\Box}}=0.48$
 the momentum-isospin entanglement is suppressed and $\mathcal{S}_{\rm vN}$ develops a pronounced dip, as shown by the black line in Fig.~\ref{fig:Entropy_d6}. We emphasize that the vanishing of the entropy occurs only at leading order in $\Delta$. At $\mathcal{O}(\Delta^2)$, contributions proportional to the forward amplitude $|m_{23}^{\rm fw}|^2$ 
  generate a small but non-zero entropy, so the dip corresponds to a sharp local minimum rather than an exact zero of the full entropy.

\subsection{Higgs basis: Product state 2}
\label{sec:higgsPS2}
We consider the initial state, which is a mixture of both the isospin singlet and triplet states, and is also inspired by the positivity condition Eq.~(\ref{App:positiv1})
\begin{equation}
|\text{in}\rangle = \frac{1}{\sqrt{V}}\,|\mathbf{p}_A \mathbf{p}_B\rangle ~(|a\rangle \otimes |b\rangle) ,
\label{eq:PS2}
 \end{equation}
where 
\bea
|a\rangle=\frac{1}{\sqrt{2}}(|1\rangle+|2\rangle), \quad |b\rangle=\frac{1}{\sqrt{2}}(|1\rangle-|2\rangle)
\eea
For this case the computational basis coefficient becomes $a_{11}=1/2,a_{12}=-1/2,a_{21}=1/2,a_{22}=-1/2$. We can calculate the von Neumann entropy as in Eq.~(\ref{eq:vNentropy} )
from two non-zero eigenvalues  $\mathcal{E}$ and $1-\mathcal{E}$  of the reduced density matrix $\tilde{\rho}$, where
\begin{equation}
   \mathcal{E}= 1-\frac{1}{4} \, \Delta \int d\Pi_2 \; (m_{11} - m_{22} + m_{23})^2+\mathcal{O}(\Delta^2) \simeq 1-\frac{1}{4} \, \Delta \lambda_I.
\end{equation}
In the above, we define 
\begin{align}
\lambda_I&=\int d\Pi_2 \; (m_{11} - m_{22} + m_{23})^2 \notag\\
&= \frac{1}{120\pi}
\Bigl[ 240\,\mathcal{C}_{H4}^2
- 120\,\frac{s}{\Lambda^2}\,\mathcal{C}_{H4}(\mathcal{C}_{H4D2} + 2\mathcal{C}_{H\Box}) \notag\\
&\quad + 20\frac{s^2}{\Lambda^4}\bigl(-2\mathcal{C}_{H4}(\mathcal{C}_1 + 3\mathcal{C}_2 + \mathcal{C}_3) 
+ (\mathcal{C}_{H4D2}+\mathcal{C}_{H\Box})^2 +  3\mathcal{C}_{H\Box}^2\bigr) \notag\\
&\quad + 5\frac{s^3}{\Lambda^6}\bigl(\mathcal{C}_{H4D2}(3\mathcal{C}_1 + 6\mathcal{C}_2 + \mathcal{C}_3) 
+ 2\mathcal{C}_{H\Box}(\mathcal{C}_1 + 6\mathcal{C}_2 + 3\mathcal{C}_3)\bigr) \notag\\
&\quad +\frac{s^4}{\Lambda^8}\bigl(3~\mathcal{C}_1^2 + \mathcal{C}_1(10\mathcal{C}_2 + \mathcal{C}_3) + 15\mathcal{C}_2^2 + 10\mathcal{C}_2\mathcal{C}_3 + 3\mathcal{C}_3^2\bigr)
\Bigr]
\end{align}

Linear Entropy for this scenario is given by
\begin{equation}
 \mathcal{S}_L(\tilde{\rho})=  \frac{1}{2} \, \Delta \int d\Pi_2 \; (m_{11} - m_{22} + m_{23})^2-  \frac{1}{2} \, \ \Delta^2   \; (m_{11} - m_{22} + m_{23})^2
\end{equation}

For the SM scenario, the entropy is controlled entirely by $\mathcal{C}_{H4}$ as $ \lambda_I\big|_{SM} = 2c_{H4}^2/\pi$. The state is never perfectly pure as $\mathcal{C}_{H4}\ne0$. As $\mathcal{C}_{H4}$ is fixed and $\Delta \le \frac{1}{16\pi}$, entropy $\mathcal{S}_{\rm vN}=0 \text{ or } 1$ is not possible.  With $\mathcal{C}_{H4} = 0.12$ and $\Delta = 0.9/(16\pi)$, we obtain
$ \mathcal{S}_{\rm vN}\big|_{SM}  \approx 7\times10^{-4}$.
At $\mathcal{O}(s/\Lambda^2)$, the interference term $- \frac{1}{\pi}\left(\frac{s}{\Lambda^2}\right)
    \mathcal{C}_{H4}(\mathcal{C}_{H4D2}+2\mathcal{C}_{H\Box})$ contributes. The entropy variation at this order is entirely 
governed by the sign of $\mathcal{C}_{H4}(\mathcal{C}_{H4D2}+2\mathcal{C}_{H\Box})$. With $\mathcal{C}_{H4} = 0.12 > 0$, 
the sign is determined purely by the combination $\mathcal{C}_{H4D2}+2\mathcal{C}_{HB}$. When 
this combination is negative, the term adds to $\lambda_I$ and entropy grows 
with energy. When positive, it subtracts and entropy decreases. When 
$\mathcal{C}_{H4D2}+2\mathcal{C}_{HB} = 0$ exactly, this entire order vanishes and the $\mathcal{O}(s^2/\Lambda^4)$
terms become the leading energy-dependent correction. At this order, both
$\mathcal{M}_{\text{d-4}}\mathcal{M}_{\text{d-8}}$ and $|\mathcal{M}_{\text{d-6}}|^2$ terms contribute.
The $|\mathcal{M}_{\text{d-6}}|^2$ term always increases $\lambda_I$ and hence entropy regardless of the signs 
of the coefficient. The  cross term is sign-indefinite, when 
$\mathcal{C}_1+3\mathcal{C}_2+\mathcal{C}_3 > 0$ it subtracts from $\lambda_I$ and competes with the 
positive definite piece, and when $\mathcal{C}_1+3\mathcal{C}_2+\mathcal{C}_3 < 0$ it reinforces 
entropy growth.
At $\mathcal{O}(s^3/\Lambda^6)$ the cross term  $\mathcal{M}_{\text{d-6}}\mathcal{M}_{\text{d-8}}$ contributes. The sign structure depends entirely on 
the relative orientation of the two sets of Wilson coefficients and therefore can either suppress or increase the entropy. 
At $\mathcal{O}(s^4/\Lambda^8)$ the term arises from $|\mathcal{M}_{\text{d-8}}|^2$ 
always increases $\lambda_I$ as it is a positive-definite term and therefore increases entropy.

For zero entropy we need  $1-\mathcal{E} = 0$, which means either $\Delta = 0$ or 
$\lambda_I = 0$. The case $\Delta = 0$ is the trivial non-interacting limit. 
The case $\lambda_I = 0$ requires a non-trivial cancellation between the Wilson coefficients. Maximum entropy requires $\mathcal{E}=1-\mathcal{E} = \frac{1}{2}$, which means $\Delta\lambda_I = 2$. The bound $\Delta < \frac{1}{16\pi}$ implies $\lambda_I> 32\pi \approx 100.5$. Since $\lambda_I$ 
is a polynomial in $s/\Lambda^2 < 1$ with coefficients
$\mathcal{C}_i\le4\pi$, the product $\Delta\lambda$ cannot reach 2 within the physical domain. 
Therefore, maximal entanglement is forbidden within the allowed parameter space of $\Delta$ and EFT coefficients. 
\begin{figure}[t]
    \centering
    \includegraphics[width=0.56\linewidth]
    {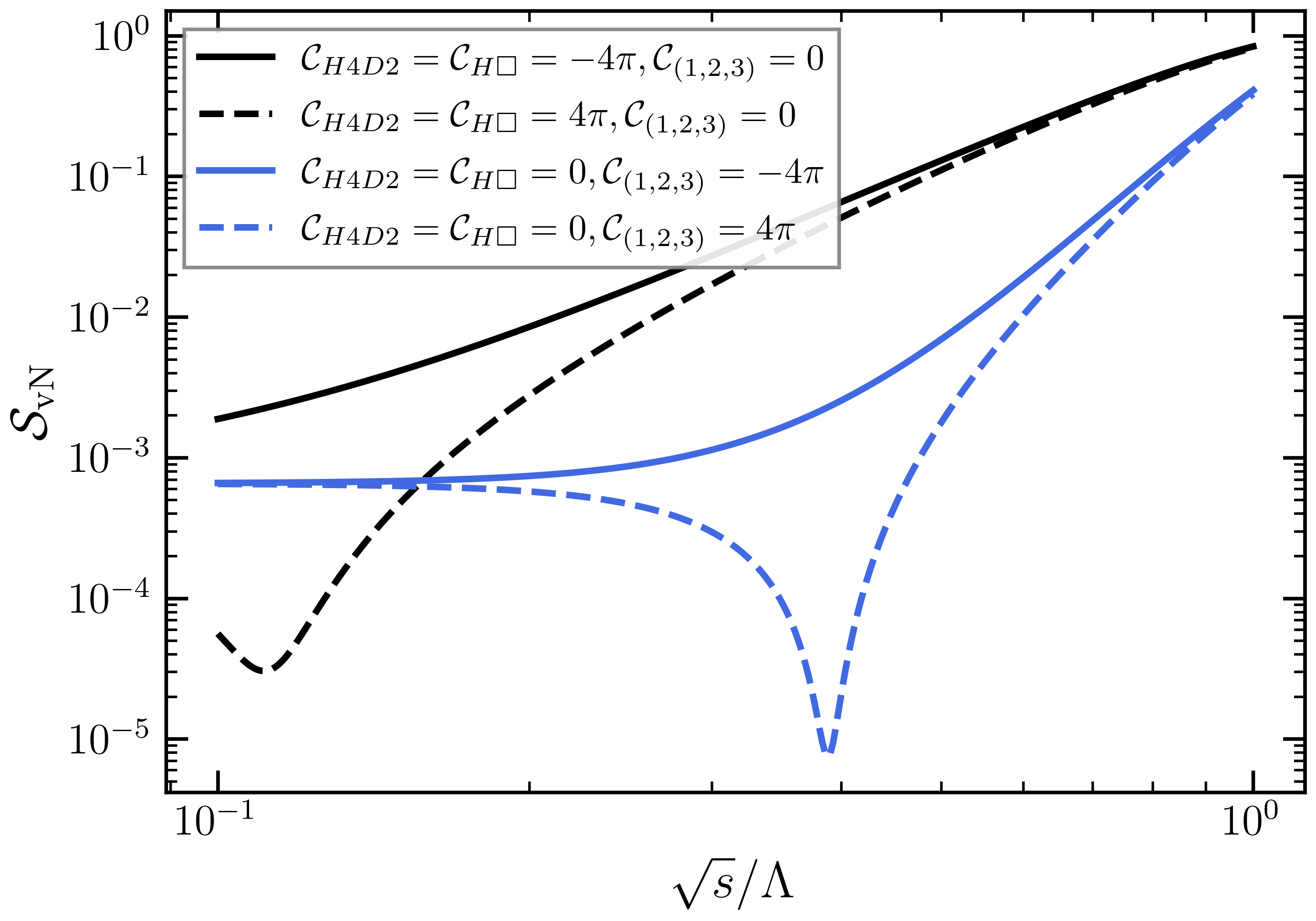}
    \caption{Variation of von Neumann entropy ($\mathcal{S}_{\rm vN}$) as a function of  $\sqrt{s}/\Lambda$. Here we fix $\Delta=0.9/16\pi$ and $\mathcal{C}_{H4}=0.12$.
    Black lines: $\mathcal{C}_{H\square}=\mathcal{C}_{H4D2}\ne0$ and $\mathcal{C}_{(1,2,3)}=0$.
    Blue lines: $\mathcal{C}_{H\square}=\mathcal{C}_{H4D2}=0$ and $\mathcal{C}_{(1,2,3)}\ne0$.}
    \label{fig:Entropy_d6_case1}
\end{figure}

  In Fig.~\ref{fig:Entropy_d6_case1}, we present the variation of $\mathcal{S}_{\rm vN}$  as a function of  $\sqrt{s}/\Lambda$ for  $\mathcal{C}_{H4}=0.12$ and $\Delta=0.9/16\pi$. The black curves are for $|\mathcal{C}_{H\square}|=|\mathcal{C}_{H4D2}|=4\pi$, $\mathcal{C}_{(1,2,3)}=0$ and blue curves are for  $\mathcal{C}_{H\square}=\mathcal{C}_{H4D2}=0$, $|\mathcal{C}_{(1,2,3)}|=4\pi$ . 
  For $\mathcal{C}_{H\square}, \mathcal{C}_{H4D2}<0$, the entropy always grows monotonically from the SM value. For $\mathcal{C}_{H\square}, \mathcal{C}_{H4D2}>0$ there occurs a cancellation due to the interference term  around $\sqrt{s}/\Lambda\sim0.2$. A similar feature is present for the case $|\mathcal{C}_{(1,2,3)}|=4\pi$. It shows the entropy can be maximal for $\sqrt{s}/\Lambda>1$, where the EFT description breaks down.
In general, the variation depends on the competition between 
the sign-indefinite interference terms and the  
positive definite squared terms. 
The squared terms drive entropy upward, whereas the interference terms are the reason for suppressing entropy.

\begin{figure}[t]
    \centering
    \includegraphics[width=0.56\linewidth]
    {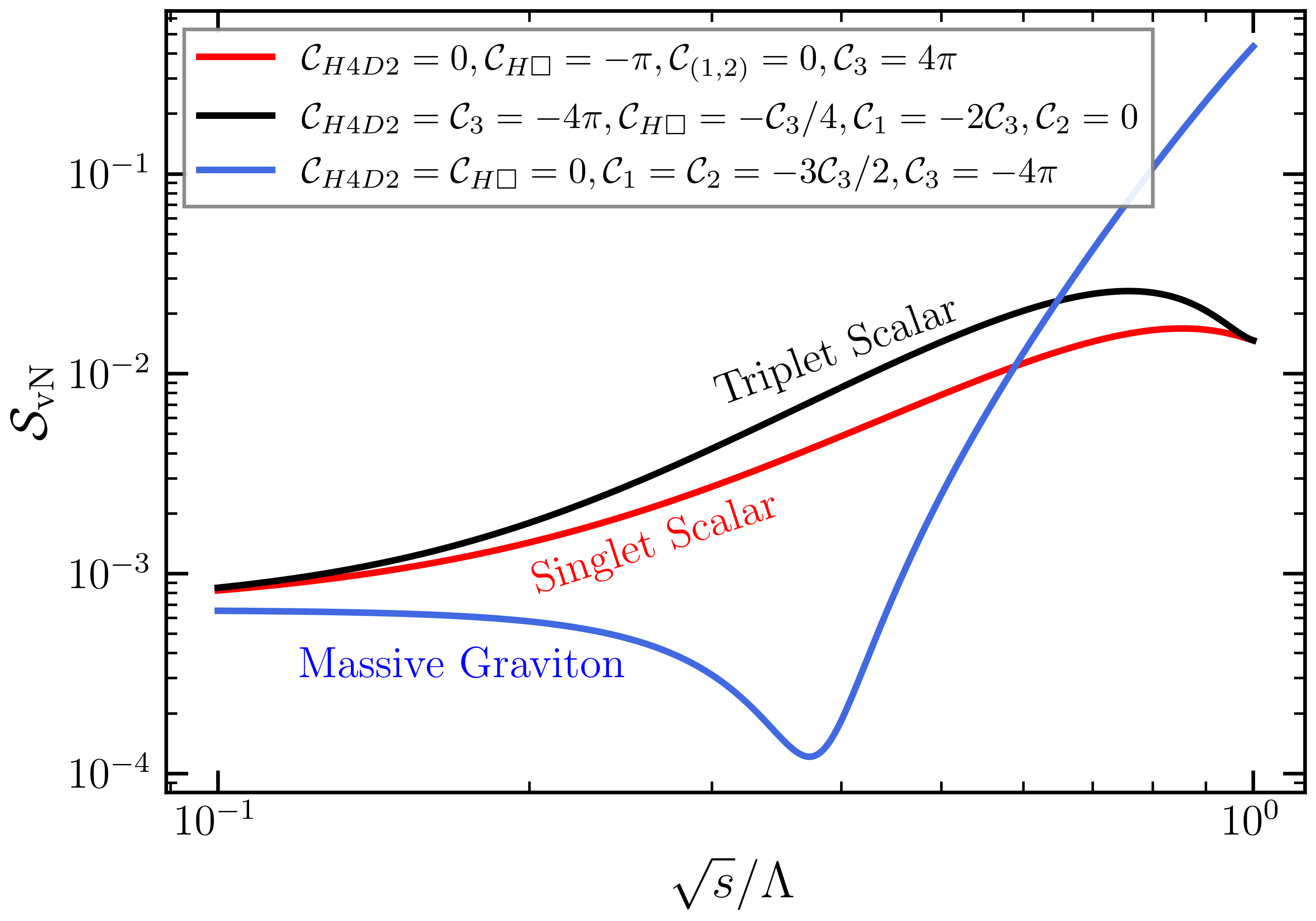}
    \caption{Variation of von Neumann entropy ($\mathcal{S}_{\rm vN}$) as a function of  $\sqrt{s}/\Lambda$. Here we fix $\Delta=0.9/16\pi$, $\mathcal{C}_{H4}=0.12$, and consider the values of the Wilson coefficient inspired by the UV complete model and positivity conditions.}
    \label{fig:Entropy_UVline}
\end{figure}

In Fig.~\ref{fig:Entropy_UVline}, we consider the values of the Wilson coefficients inspired by the UV complete model and positivity conditions. Here we fix $\Delta=0.9/16\pi$, $\mathcal{C}_{H4}=0.12$. The red curve shows variation of von Neumann entropy ($\mathcal{S}_{\rm vN}$)  for the choice of coupling $\mathcal{C}_{H4D2}=\mathcal{C}_1=\mathcal{C}_2=0$, $\mathcal{C}_{H\Box}=-\pi$ and $\mathcal{C}_3=4\pi$, which is inspired by  Singlet scalar model (see Eq.~(\ref{eq:UV_singlet})). It is crucial to note that only $\mathcal{C}_3\ge0$ is allowed by positivity conditions for the singlet scalar model.
As discussed earlier, $|\mathcal{M}_{\text{d-4,6,8}}|^2$  terms always increase $\lambda_I$ and hence entropy regardless of the signs of the coefficients. In this case, the interference terms are either sub-leading or add constructively with the $|\mathcal{M}_{\text{d-4,6,8}}|^2$ terms. Therefore, the entropy grows monotonously with energy.  Similarly, the black curve is for the triplet scalar model, where we fix the coupling $\mathcal{C}_{H4D2}=\mathcal{C}_3=-4\pi$, $\mathcal{C}_{H\Box}=-\mathcal{C}_3/4$, $\mathcal{C}_1=-2\mathcal{C}_3$ and $\mathcal{C}_2=0$. The positivity conditions allow either $\mathcal{C}_1\ge0$ (or $\mathcal{C}_3\le0$). Similar to the Singlet Scalar model, here also entropy grows monotonously with $\sqrt{s}/\Lambda$.

The blue curve in Fig.~\ref{fig:Entropy_UVline} is for the choice of coupling $\mathcal{C}_{H4D2}=\mathcal{C}_{H\Box}=0$, $\mathcal{C}_1=\mathcal{C}_2=-3\mathcal{C}_3/2$ and $\mathcal{C}_3=-4\pi$, which is inspired by  Massive Graviton model (see Eq.~(\ref{eq:UV_graviton})).
The positivity conditions force $\mathcal{C}_3\le0$.
As discussed earlier, $|\mathcal{M}_{\text{d-4}}|^2$ and $|\mathcal{M}_{\text{d-8}}|^2$ terms always increase $\lambda_I$ and hence entropy regardless of the signs 
of the coefficient. However, as the cross term  $\mathcal{M}_{\text{d-4}}\mathcal{M}_{\text{d-8}}$ is negative for $\mathcal{C}_3\le0$, it partially cancels with the $|\mathcal{M}_{\text{d-4}}|^2$ and $|\mathcal{M}_{\text{d-8}}|^2$ term when energy is  low. As energy increases the $|\mathcal{M}_{\text{d-8}}|^2$  term dominates and enhances the entropy. 

\begin{figure}[h]
    \centering
    \includegraphics[width=0.456\linewidth]
    {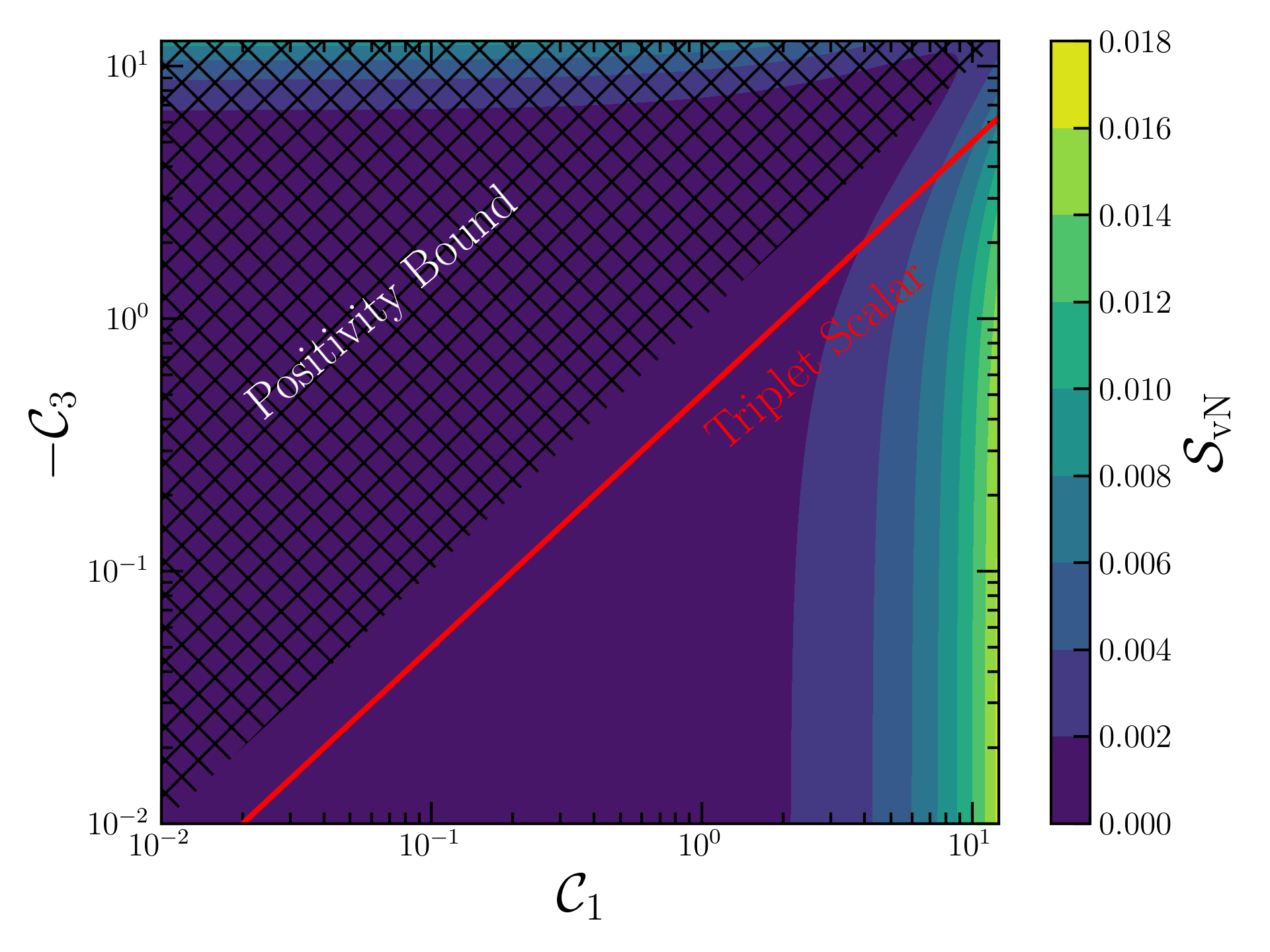}
    \includegraphics[width=0.456\linewidth]
    {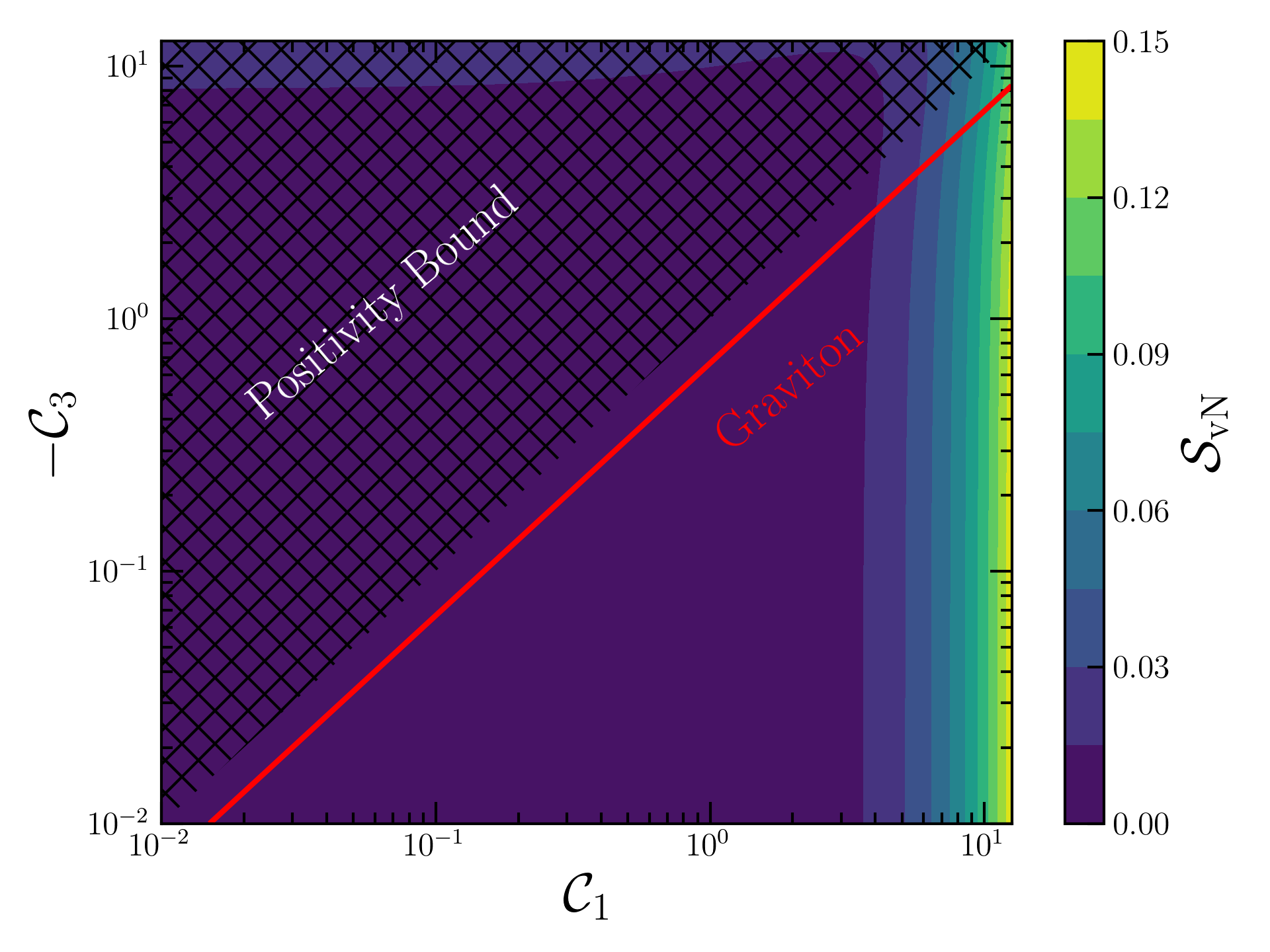}
    \caption{Variation of von Neumann entropy ($\mathcal{S}_{\rm vN}$) on $\mathcal{C}_1 \text{ vs } \mathcal{C}_3$  plane. Here we fix $\Delta=0.9/16\pi$ and $\sqrt{s}/\Lambda=0.9$.
    Left: Triplet scalar,  
    Right: Graviton.}
    \label{fig:Entropy_2Dcase1}
\end{figure}

In Fig.~\ref{fig:Entropy_2Dcase1}, we plot variation of von Neumann entropy ($\mathcal{S}_{\rm vN}$) on $\mathcal{C}_1 \text{ vs } \mathcal{C}_3$  plane for fixed $\Delta=0.9/16\pi$ and $\sqrt{s}/\Lambda=0.9$. Here we consider the coupling choices as discussed for the graviton and triplet scalar UV models. The red line of the left plot corresponds to the Triplet scalar. The red line of the right plot corresponds to the massive Graviton. The hatched region is excluded from the positivity bound: $\mathcal{C}_1+\mathcal{C}_2\geq 0, ~\mathcal{C}_2 \geq 0,~\mathcal{C}_1+\mathcal{C}_2+\mathcal{C}_3\geq 0$ discussed in Appendix.~\ref{App:positivity}.

\subsection{Higgs basis: Bell states}
Here we consider maximally entangled Bell states in isospin space but separable across momentum-isospin space
\bea
|\text{in}\rangle_T = \frac{1}{\sqrt{V}}\,|\mathbf{p}_A \mathbf{p}_B\rangle \frac{1}{\sqrt{2}}\,~(|11\rangle + |22\rangle), \nonumber \\
|\text{in}\rangle_S = \frac{1}{\sqrt{V}}\,|\mathbf{p}_A \mathbf{p}_B\rangle \frac{1}{\sqrt{2}}\,~(|12\rangle -|21\rangle).
 \eea
 In the above equation, $|\text{in}\rangle_S$ ($|\text{in}\rangle_T$) is an isospin singlet (triplet)  state. Therefore, the scattering does not change the initial entanglement.
For the Bell states the momentum reduced density matrix $\tilde{\rho}$ has only one non-zero eigenvalue, which implies the scattering process does not entangle momentum and isospin space. 
\subsection{General Higgs basis: Product state 1}
For this case, we consider an initial isospin state inspired by the positivity condition Eq.~(\ref{App:positiv2}),

\begin{equation}
|\text{in}\rangle = \frac{1}{\sqrt{V}}\,|\mathbf{p}_A \mathbf{p}_B\rangle ~(|a\rangle \otimes |b\rangle) ,
\label{eq:GHB_PS1}
 \end{equation}
where,
\bea
|a\rangle=\frac{1}{\sqrt{2}}(|H_1\rangle+|H_1^*\rangle) \equiv\frac{1}{\sqrt{2}}(|1\rangle+|2\rangle),\nonumber\\ 
|b\rangle=\frac{1}{\sqrt{2}}(|H_1\rangle-|H_1^*\rangle) \equiv\frac{1}{\sqrt{2}}(|1\rangle-|2\rangle),
\eea
with the computational  basis coefficients  $a_{11}=1/2,a_{12}=-1/2,a_{21}=1/2,a_{22}=-1/2$.
Note that the initial state is similar to the state considered in section.~\ref{sec:higgsPS2} however, the scattering is  $H_\alpha \tilde{H}_{\beta}\to H_\gamma \tilde{H}_{\delta}$. Therefore, the reduced density matrix has the same form as in section.~\ref{sec:higgsPS2} but the scattering amplitudes are from Eq.~(\ref{eq:AmpHHc}). The two non-zero eigenvalues of $\tilde{\rho}$ are  $\mathcal{E}$ and $1-\mathcal{E}$, where
\begin{equation}
   \mathcal{E}= 1-\frac{1}{4} \, \Delta \int d\Pi_2 \; (\hat m_{11} - \hat m_{22} + \hat m_{23})^2+\mathcal{O}(\Delta^2)\simeq 1-\frac{1}{4} \, \Delta \lambda_I.
\end{equation}
In the above equation  
\begin{align}
\lambda_I&=\int d\Pi_2 \; (\hat m_{11} - \hat m_{22} + \hat m_{23})^2\notag\\
&= \frac{1}{120\pi}
\Bigl[ 240\,\mathcal{C}_{H4}^2
- 120\,\frac{s}{\Lambda^2}\,\mathcal{C}_{H4}(\mathcal{C}_{H4D2} - 4~\mathcal{C}_{H\Box}) \notag\\
&\quad + 20\frac{s^2}{\Lambda^4}\bigl(-2\mathcal{C}_{H4}(\mathcal{C}_1 + \mathcal{C}_2 + 3~\mathcal{C}_3) 
+ (\mathcal{C}_{H4D2}-3\mathcal{C}_{H\Box})^2 +  3\mathcal{C}_{H\Box}^2\bigr) \notag\\
&\quad + 5\frac{s^3}{\Lambda^6}\bigl(\mathcal{C}_{H4D2}(3\mathcal{C}_1 + \mathcal{C}_2 + 6\mathcal{C}_3) 
-8  \mathcal{C}_{H\Box}(\mathcal{C}_1 + \mathcal{C}_2 + 3\mathcal{C}_3)\bigr) \notag\\
&\quad +\frac{s^4}{\Lambda^8}\bigl(3~\mathcal{C}_1^2 + 10 \mathcal{C}_3(\mathcal{C}_1 + \mathcal{C}_2) +  \mathcal{C}_1\mathcal{C}_2 +3\mathcal{C}_2^2+ 15\mathcal{C}_3^2\bigr)
\Bigr].
\end{align}
\subsection{General Higgs basis: Product state 2}
For this case, we consider the state inspired by the positivity condition Eq.~(\ref{App:positiv3}), with $a_{11}=1/2,a_{12}=1/2,a_{21}=1/2,a_{22}=1/2$, namely,
\bea
|a\rangle=|b\rangle=\frac{1}{\sqrt{2}}(|H_1\rangle+|H_1^*\rangle)\equiv\frac{1}{\sqrt{2}}(|1\rangle+|2\rangle).
\label{eq:GPS2}
\eea
This is a triplet isospin state based on Eq.~(\ref{Eq:CtoI}), therefore, scattering will not change the initial entanglement entropy. This is evident if we calculate  $\tilde{\rho}$, which has only one non-zero eigenvalue. As the amplitudes satisfy the relation  $\hat m_{11} = \hat m_{22} + \hat m_{23}$ the eigenvalue $\mathcal{E}$\footnote{Note that $\mathcal{O}(\Delta^2)$ terms also vanish when $\hat m_{11} = \hat m_{22} + \hat m_{23}$.  } 
\begin{equation}
   \mathcal{E}= 1-\frac{1}{4} \, \Delta \int d\Pi_2 \; (\hat m_{11} - \hat m_{22} - \hat m_{23})^2+\mathcal{O}(\Delta^2)=1.
\end{equation}
\section{Higgs scattering and concurrence} \label{sec:Concurence}
In this section, we investigate the entanglement between the two isospin qubits. To this end, the momentum degrees of freedom in the final state must be traced out, a procedure that was carried out in the previous section. We need different entanglement measures depending on whether the final state isospin system is pure or mixed. The momentum reduced density matrix $\tilde{\rho}$ represents a mixed state for $\Delta<\Delta_{\rm max}$. For such a mixed state, we calculate concurrence from the momentum reduced density matrix as a measure of entanglement between two isospin qubits. For $\Delta=\Delta_{\rm max}$, the isospin state is pure, so one can calculate concurrence right from the state vector, which we discuss in the following subsection.

Following the same logic as in the case of isospin-momentum entanglement, the initial entanglement between the two isospin qubits remains unchanged by the scattering process whenever the initial state is an isospin singlet or triplet state.

The concurrence for the reduced density matrix $\tilde{\rho}$ up to leading order in $\Delta$ is given by~\cite{Kowalska:2025qmf}
\be
C(\tilde{\rho})=2\Delta\left|\sum_{\alpha,\beta,\gamma,\delta}(-1)^{\alpha+\beta}a_{\alpha\beta}a_{\gamma\delta}(-\mathcal{M}_{\rm fw})_{\bar{\alpha}\bar{\beta},\gamma\delta}\right|, \quad \text{where }\bar{\alpha}=3-\alpha.
\ee
For the amplitude  matrix defined in Eq.~(\ref{eq:Mmatrix}), we get
\bea
C(\tilde{\rho})=
2\Delta\left| -2\,m_{11}\, a_{11} \,a_{22}+2\,m_{22}\, a_{21}\, a_{12}+m_{23} \,a_{21}^2+m_{32} \,a_{12}^2 \right|.
\eea
For the initial state,  $|\text{in}\rangle = \frac{1}{\sqrt{V}}\,|\mathbf{p}_A \mathbf{p}_B\rangle |12\rangle$,
 we obtain
 $C(\tilde{\rho})\simeq2\Delta|m_{32}^{\rm fw}|$.
 For the initial state in Eq.~(\ref{eq:PS2}), where $a_{11}=1/2,a_{12}=-1/2,a_{21}=1/2,a_{22}=-1/2$ , $C(\tilde{\rho}) = \Delta  \left| m_{11}-m_{22}+m_{23}\right|$. For the initial state in Eq.~(\ref{eq:GPS2}), where all $a_{\alpha\beta}=1/2$,  $C(\tilde{\rho}) = \Delta  \left| \hat m_{11}-\hat m_{22}-\hat m_{23}\right|=0$. As stated before, this initial state is a triplet isospin state; therefore, the concurrence of the final state remains the same as in the initial state.
\subsection{General initial states and concurrence}
\label{subsec:GenInconc}
 Here, we discuss entanglement when the final isospin state is pure. As mentioned above for $\Delta=\Delta_{\rm max}$, the isospin state is pure. Therefore, in this case, entanglement can be characterized by standard pure state measures.
We consider a general pure initial state as a tensor product of two general single particle states:
\bea
|{\rm in}\rangle = (x_1 |1\rangle+x_2|2\rangle)\otimes  (y_1 |1\rangle+y_2|2\rangle)\equiv |a\rangle \otimes |b\rangle, \label{generalstate}
\eea
with $|x_1|^2+|x_2|^2=|y_1|^2+|y_2|^2=1$.  Then, after the Higgs scattering, we can obtain the final state  in terms of the scattering amplitudes, as follows~\cite{Carena:2023vjc},
\bea
|{\rm out}\rangle =(I+i \mathcal{M})|{\rm in}\rangle= c_{ij}|i\rangle \otimes |j\rangle=c_{ij}|ij\rangle, \quad i,j=1, 2,
\eea
where
\begin{align}
c_{11} &= (1 + i \mathcal{M}_{11,11}) x_1 y_1
+ i \mathcal{M}_{12,11} x_1 y_2 
+ i \mathcal{M}_{21,11} x_2 y_1 
+ i \mathcal{M}_{22,11} x_2 y_2  \nonumber  \\
&=  (1 + i \mathcal{M}_{11,11}) x_1 y_1, \\
c_{12} &= i \mathcal{M}_{11,12} x_1 y_1 
+ (1 + i \mathcal{M}_{12,12})  x_1 y_2 
+ i \mathcal{M}_{21,12} x_2 y_1 
+ i \mathcal{M}_{22,12} x_2 y_2 \nonumber  \\
&= (1 + i \mathcal{M}_{12,12})  x_1 y_2 
+ i \mathcal{M}_{21,12} x_2 y_1, \\
c_{21} &= i \mathcal{M}_{11,21}x_1 y_1 
+ i \mathcal{M}_{12,21}  x_1 y_2 
+ (1 + i \mathcal{M}_{21,21}) x_2 y_1 
+ i \mathcal{M}_{22,21} x_2 y_2 \nonumber \\
&= 
i \mathcal{M}_{12,21}  x_1 y_2 
+ (1 + i \mathcal{M}_{21,21}) x_2 y_1, \\
c_{22} &= i \mathcal{M}_{11,22} x_1 y_1 
+ i \mathcal{M}_{12,22}  x_1 y_2 
+ i \mathcal{M}_{21,22} x_2 y_1 
+ (1 + i \mathcal{M}_{22,22}) x_2 y_2 \nonumber \\
&=  (1 + i \mathcal{M}_{22,22}) x_2 y_2.
\end{align}
Then, the concurrence $C(|{\rm out}\rangle) =2 |c_{11}c_{22} - c_{12}c_{21}|$ reads
\begin{align}
C(|{\rm out}\rangle)
&=2|\mathcal{M}_{12,21}(- i + \mathcal{M}_{12,12}) (x_1y_2-x_2 y_1)^2|
\end{align}
Here, we used $\mathcal{M}_{11,11}=\mathcal{M}_{22,22}$ and $\mathcal{M}_{11,11}-\mathcal{M}_{12,12}=\mathcal{M}_{12,21}$. Then, for arbitrary $x_1, x_2, y_1, y_2$,
setting $C(|out\rangle)=0$ leads to the conditions $\mathcal{M}_{12,21}=(\mathcal{M}_1-\mathcal{M}_0)/2=0$.



In the following, we discuss concurrence for the scattering of the SM Higgs and its complex conjugate. 
We take an alternative basis (as in section~\ref{sec:AlterBasis}) for the general initial states, $\{|H_1\rangle,|H^*_1\rangle\}$,  motivated by the positivity bounds in $(2)$ and $(3)$, as follows, 
\bea
|{\rm in}\rangle = (x_1 |H_1\rangle+z_1|H^*_1\rangle)\otimes  (y_1 |H_1\rangle+w_1|H^*_1\rangle) \equiv (x_1 |1\rangle+z_1|2\rangle)\otimes  (y_1 |1\rangle+w_1|2\rangle) \quad
\eea
with $|x_1|^2+|z_1|^2=|y_1|^2+|w_1|^2=1$.  Then, after the Higgs scattering, we obtain the final state  in terms of the scattering amplitudes, as follows,
\bea
|{\rm out}\rangle =(I+i  \mathcal{\hat M})|{\rm in}\rangle= d_{ij}|i\rangle \otimes |j\rangle=d_{ij}|ij\rangle, \quad i,j=1, 2.
\eea
where the coefficients  $d_{ij}$ are
\begin{align}
d_{11} & =(1 + i \mathcal{\hat M}_{11,11}) x_1 y_1, \\
d_{12} &= (1 + i \mathcal{\hat M}_{12,12})  x_1 w_1 
+ i \mathcal{\hat M}_{21,12}  z_1 y_1, \\
d_{21} &= i \mathcal{\hat M}_{12,21}  x_1 w_1 
+ (1 + i \mathcal{\hat M}_{21,21}) z_1 y_1, \\
d_{22} &= (1 + i \mathcal{\hat M}_{22,22}) z_1 w_1.
\end{align}
We can simplify the concurrence  $C(|{\rm out}\rangle) = 2|d_{11}d_{22} - d_{12}d_{21}|$ using the relations $\mathcal{\hat M}_{11,11}=\mathcal{\hat M}_{22,22}$ and $\mathcal{\hat M}_{11,11}-\mathcal{\hat M}_{12,12}=\mathcal{\hat M}_{12,21}$ as
\begin{align}
C(|{\rm out}\rangle)
&=2\left| \mathcal{\hat M}_{12,21}\left( -i+ \mathcal{\hat M}_{12,12}\right)  (x_1 w_1-z_1 y_1)^2 \right|
\end{align}

\subsection{Concurrence for momentum measured final state}
To quantify the entanglement between isospins, one can also measure the final state momenta instead of tracing out. This corresponds to a detector registering both outgoing particle momenta at a fixed scattering angle.
After a measurement of the final state momenta around $\mathbf{p}_C$, $\mathbf{p}_D$ is performed, the final state $|\mathrm{out}\rangle$ is projected onto the measured momentum state $|f\rangle \approx \frac{1}{\sqrt{V}}|\mathbf{p}_C\mathbf{p}_D\rangle$. The final isospin state becomes pure after momentum measurement, i.e., $\mathrm{Tr}(\hat{\rho}^2)=1$ for the corresponding density matrix .
 The entanglement in the two-qubit flavor space can then be quantified by concurrence as follows.

The density matrix $\hat{\rho}$ in the $|\alpha\beta\rangle\langle\gamma\delta|$ basis in terms of scattering amplitude $\mathcal{M}_{\gamma\delta,\alpha\beta}$ evaluated at the measured momenta is given by~\cite{Kowalska:2024kbs}
\begin{equation}
  \hat{\rho}_{\alpha\beta,\gamma\delta}
  = \frac{A_{\alpha\beta}A_{\gamma\delta}^*}
    {\sum_{\epsilon\rho}|A_{\epsilon\rho}|^2},
  \qquad
  A_{\gamma\delta} = \sum_{\alpha\beta}
  \mathcal{M}_{\gamma\delta,\alpha\beta}\,a_{\alpha\beta},
  \label{eq:rho_hat}
\end{equation}
 and the  two-qubit entanglement is
quantified either by the concurrence (or by entropy, as for a  pure state both are related).
\begin{equation}
  C(\hat\rho) = 2 \frac{\left|(\sum_{\alpha\beta}
  \mathcal{M}_{11,\alpha\beta}\,a_{\alpha\beta}) (\sum_{\alpha\beta}
  \mathcal{M}_{22,\alpha\beta}\,a_{\alpha\beta})-(\sum_{\alpha\beta}
  \mathcal{M}_{12,\alpha\beta}\,a_{\alpha\beta})(\sum_{\alpha\beta}
  \mathcal{M}_{21,\alpha\beta}\,a_{\alpha\beta})\right|}{\sum_{\alpha\beta}|\sum_{\gamma\delta}
\mathcal{M}_{\alpha\beta,\gamma\delta}\,a_{\gamma\delta}|^2}
  \label{eq:concurrence}
\end{equation}
Since the scattering amplitude $\mathcal{M}_{\gamma\delta,\alpha\beta}$ depends on the scattering angle, the concurrence also exhibits a non-trivial angular dependence
More explicitly, for the amplitude matrix $\mathcal{M}({H_\alpha H_\beta \to H_\gamma H_\delta})$ in Eq.~(\ref{eq:scat01}) we get
\be
C(\hat\rho) = \frac{E}{F},
\ee
with
\bea
E&=&2\left|m_{11}^2 a_{11} a_{22}-\left(m_{22}\, a_{21}+m_{23} \,a_{12}\right) \left(m_{22}\, a_{12}+m_{23}\, a_{21}\right)\right|, \\
F&=&|m_{11}|^2 \left(\left| a_{11}\right| {}^2+\left|a_{22}\right| {}^2\right)+\left(\left| a_{12}\right| {}^2+\left| a_{21}\right| {}^2\right) \left(| m_{22}|^2+| m_{23}|^2\right) \nonumber \\
&&+\left(a_{12} a_{21}^*+a_{21} a_{12}^*\right) \left(m_{23} m_{22}^*+m_{22} m_{23}^*\right).
\eea
For the initial state as in Eq.~(\ref{eq:PS2}), it is simplified to 
\be
C(\hat\rho) =\frac{|-m_{11}^2+(m_{22}-m_{23})^2|}{2(|m_{11}|^2+(m_{22}-m_{23})(m_{22}^*-m_{23}^*))}\,.
\ee
Similarly in $H_\alpha \tilde{H}_{\beta}\to H_\gamma \tilde{H}_{\delta}$ scattering for the initial state Eq.~(\ref{eq:GPS2}), we obtain
\be
C(\hat\rho) =\frac{|\hat m_{11}^2-(\hat m_{22}+\hat m_{23})^2|}{2(|\hat m_{11}|^2+(\hat m_{22}+\hat m_{23})(\hat m_{22}^*+\hat m_{23}^*))}.
\label{eq:Mmom}
\ee

As discussed earlier, scattering can induce non-zero $C(\hat\rho)$ for the initial state in Eq.~(\ref{eq:PS2}) as it is an admixture of singlet and triplet isospin states. Whereas the initial state in Eq.~(\ref{eq:GPS2}) is a triplet isospin state, scattering does not change the 
initial entanglement. This is evident from  Eq.~(\ref{eq:Mmom}), as $C(\hat\rho)=0$ for $\hat m_{11}=\hat m_{22}+\hat m_{23}$.
\section{Positivity bounds and entanglement} \label{sec:positivity}

We consider general initially separable states for Higgs scattering and discuss the positivity bounds and the concurrence as well as entanglement power.

\subsection{Positivity bounds}

For the general initial states in the Higgs doublet basis, we also consider the positivity bounds. 
Using the general results on the positivity bounds for the most general superposition of states in Appendix.~\ref{App:positivity}, we restrict ourselves to the Higgs doublet basis, not its complex conjugate, so we can impose the positivity bounds
for the elastic scattering, $ab\to ab$ by
\bea
{\overline M}(ab\to ab)\equiv \frac{1}{2} \frac{\partial^2}{\partial s^2} M(ab\to ab)(s,t=0)\geq 0, \label{positivity1}
\eea
with
\bea
M(ab\to ab)=x_\alpha y_\beta x^*_\gamma y^*_\delta M_{\alpha\beta,\gamma\delta}.
\eea
Then, taking $ x_\alpha,  y_\alpha$ as being real, we express  ${\overline M}(ab\to ab)$ in Eq.~(\ref{positivity1}) as
\bea
{\overline M}(ab\to ab)=\Big(\mathcal{C}_1+\mathcal{C}_2\Big) \frac{1}{2}(x_2 y_1-x_1 y_2)^2\geq 0.\label{superampl1}
\eea
Therefore, for arbitrary $x_1, x_2, y_1, y_2$, we obtain the positivity bounds from the superposed states in the Higgs doublet basis \cite{Bi:2019phv,Remmen:2019cyz,Zhang:2021eeo,Yamashita:2020gtt,Kim:2023pwf,Kim:2023bbs}  as
\bea
\mathcal{C}_1+\mathcal{C}_2\geq 0,
\eea
which is the case $(1)$ in Appendix.~\ref{App:positivity}.
We note that the other positivity conditions can be derived from the superposed states of the Higgs fields and its complex conjugates, as shown for case $(2)$ and $(3)$ in Appendix.~\ref{App:positivity}.

\subsection{Entanglement suppression/generation}

Now we focus on  dimension-8 operators for Higgs scattering and derive the conditions for isospin entanglement suppression/generation for a generic separable state $|{\rm in}\rangle = (x_1 |1\rangle+x_2|2\rangle)\otimes  (y_1 |1\rangle+y_2|2\rangle)$ using concurrence measure. Moreover, to connect these conditions with the positivity bounds we assume forward ($t\to 0$, $u\to -s$) and backward ($u\to 0$, $t\to -s$)  scattering  limit. In the forward scattering  limit ($t\to0,u\to-s$) the relevant amplitudes for $HH \to HH$  are
\be
   \mathcal{M}_1=m_{11}= (\mathcal{C}_1+2~\mathcal{C}_2+\mathcal{C}_3 ) \frac{s^2}{2\Lambda ^4},
\ee
\be
 \mathcal{M}_0= m_{22}-m_{23}= (\mathcal{C}_1-\mathcal{C}_3) \frac{s^2}{2\Lambda ^4}, 
        \ee  
        \be
   m_{23}= (\mathcal{C}_2+\mathcal{C}_3) \frac{s^2}{2\Lambda ^4} .
\ee
 and  the forward  scattering amplitudes for  
$H\tilde{H} \to H\tilde{H}$ are
\be
\mathcal{M}_1=\hat m_{11}= (\mathcal{C}_1+\mathcal{C}_2 ) \frac{s^2}{2\Lambda ^4},
\ee
\be
\mathcal{M}_0=\hat m_{22}-\hat m_{23}= (\mathcal{C}_1+3~\mathcal{C}_2+2~\mathcal{C}_3 ) \frac{s^2}{2\Lambda ^4},
\ee

\be
\hat m_{23}= -(\mathcal{C}_2+\mathcal{C}_3 ) \frac{s^2}{2\Lambda ^4}.
\ee

\begin{table}[t]
\centering
\setlength{\tabcolsep}{4pt}
\renewcommand{\arraystretch}{1.4}
{\small
\begin{tabular}{l l l l l}
\hline \hline
Channel & Kinematic limit & Condition & Mechanism & Entanglement (UV model) \\
\hline
\multirow{6}{*}{$HH \to HH$} 
 &  Forward  
   & $\mathcal{C}_1 + 2\mathcal{C}_2 + \mathcal{C}_3 = 0$ 
   & $\mathcal{M}_1 = 0$ & Entangled \\
 & 
   & $\mathcal{C}_1 - \mathcal{C}_3 = 0$ 
   & $\mathcal{M}_0 = 0$ & Entangled \\
 & 
   & $\mathcal{C}_2 + \mathcal{C}_3 = 0$ 
   & $\mathcal{M}_0 = \mathcal{M}_1$ & Separable\\
\cline{2-5}
 & Backward 
   & $\mathcal{C}_1 + 2\mathcal{C}_2 + \mathcal{C}_3 = 0$ 
   & $\mathcal{M}_1 = 0$& Entangled \\
 & 
   & $\mathcal{C}_1 - \mathcal{C}_3 = 0$ 
   & $\mathcal{M}_0 = 0$& Entangled\\
 & 
   & $\mathcal{C}_1 + \mathcal{C}_2 = 0$ 
   & $\mathcal{M}_0 = \mathcal{M}_1$ & Separable (Singlet Scalar) \\
\hline
\multirow{4}{*}{$H\tilde{H} \to H\tilde{H}$} 
 & Forward 
   & $\mathcal{C}_1 + \mathcal{C}_2 = 0$ 
   & ${\mathcal{M}}_1 = 0$ & Entangled (Singlet Scalar)\\
 & 
   & $\mathcal{C}_1 + 3\mathcal{C}_2 + 2\mathcal{C}_3 = 0$ 
   & ${\mathcal{M}}_0 = 0$ & Entangled (Triplet scalar) \\
 & 
   & $\mathcal{C}_2 + \mathcal{C}_3 = 0$ 
   & ${\mathcal{M}}_0 = {\mathcal{M}}_1$ & Separable\\
\cline{2-5}
 & Backward 
   & $\mathcal{C}_1 + \mathcal{C}_3 = 0$ 
   & ${\mathcal{M}}_0 = {\mathcal{M}}_1 = 0$ & Separable \\
\hline \hline
\end{tabular}
}
\caption{Entanglement conditions from dimension-8 operators in the 
forward and backward scattering limits, for both the $HH \to HH$ and 
$H\tilde{H} \to H\tilde{H}$ channels. Each condition corresponds to one of 
three mechanisms: the singlet amplitude vanishes ($\mathcal{M}_0 = 0$), 
the triplet amplitude vanishes ($\mathcal{M}_1 = 0$), or the two amplitudes 
become equal ($\mathcal{M}_0 = \mathcal{M}_1$, equivalently the off-diagonal 
$m_{23}$ or $\hat m_{23}$ vanishes). We note that some of the above entanglement conditions are satisfied in the UV complete models (Eq.~(\ref{eq:UV_singlet}),Eq.~(\ref{eq:UV_triplet})), which are mentioned in the last column. }
\label{tab:suppression}
\end{table}

In Table~\ref{tab:suppression},  using the concurrence introduced for initial product states in Sec.~\ref{subsec:GenInconc}, we present entanglement generation or suppression conditions from dimension-8 operators in the 
forward and backward scattering limits, for both the $HH \to HH$ and 
$H\tilde{H} \to H\tilde{H}$ channels. The table reveals a few patterns worth noting. First, in the $HH \to HH$ 
channel, the conditions that make the triplet or singlet amplitude vanish, 
$\mathcal{C}_1 + 2\mathcal{C}_2 + \mathcal{C}_3 = 0$ and $\mathcal{C}_1 = 
\mathcal{C}_3$, are the same in both the forward and backward limits. The same relation works in two different kinematic regimes. The outgoing state can be entangled when either singlet or triplet amplitude vanishes. Second, the condition 
$\mathcal{C}_2 + \mathcal{C}_3 = 0$, which makes the singlet and triplet 
amplitudes equal, appears in the forward limit of both the $HH \to HH$ and 
$H\tilde{H} \to H\tilde{H}$ channels. A single relation among the Wilson 
coefficients therefore suppresses entanglement in two different scattering  
processes at once. Third, the backward limit of $H\tilde{H} 
\to H\tilde{H}$ behaves differently from the others. The condition 
$\mathcal{C}_1 + \mathcal{C}_3 = 0$ makes both amplitudes vanish simultaneously, which means there is no scattering at all rather than entanglement suppression. 
Finally, the condition $\mathcal{C}_1 + \mathcal{C}_2 = 0$, for which the dimension-8 Wilson coefficients sit on the boundary of the positivity allowed region, suppresses entanglement  in the backward (forward) limit for  $HH \to HH$ 
$(H\tilde{H} \to H\tilde{H})$.

It is worthwhile to make a few comments on UV complete models for which entanglement suppression is relevant.
For instance, for a singlet scalar or radion (in Eq.~(\ref{eq:UV_singlet})), we have $\mathcal{C}_1=\mathcal{C}_2=0$, so the entanglement is suppressed in the backward limit for $HH$ scattering (due to  ${\cal M}_0={\cal M}_1$)
For a triplet scalar (see Eq.~(\ref{eq:UV_triplet})),  we have $\mathcal{C}_2=0$ and  $\mathcal{C}_1=-2\mathcal{C}_3$, so  the entanglement can be generated  in the forward limit for $H {\tilde H}$ scattering.

\subsection{Entanglement power}

 While our previous discussions mainly focus on entanglement measures that depend on the particular choices of the initial state or the kinematics (namely, forward or backward limits), we also briefly comment on the entanglement power of the S-matrix. Entanglement power is defined as the average entanglement generated over all possible initial product states. We study the entanglement power in Higgs scattering and discuss the resulting entanglement suppression in relation to our discussion based on concurrence in the previous subsection.

Writing the isospin singlet and triplet scattering matrix in terms of the phase shift, $S_I=e^{i2\delta_I}$ with $I=0,1$, and using the leading order approximation $S_I=1+i2\delta_I$, we have $\mathcal{M}_I=-i(e^{i2\delta_I}-1)$.  Applying the spin entanglement in nucleon scattering~\cite{Beane:2018oxh} to the Higgs isospin states in our case, the entanglement power of the S-matrix is given by 
\begin{equation}
\mathcal{E}( S)=\frac16\sin^2(2(\delta_1-\delta_0)). \end{equation}
where the S-matrix has the form
\begin{equation}
 S= \frac{1}{2}\left(e^{2i\delta_1}+e^{2i\delta_0}\right)\hat I+\frac{1}{2}\left(e^{2i\delta_1}-e^{2i\delta_0}\right)\mathrm{SWAP}.
\end{equation}
The entanglement power vanishes at four points $\delta_1-\delta_0=m\pi/2$.
At these points the S-matrix is either the Identity or the SWAP operator i.e. $S\propto I$ or $S \propto(I+\sigma.\sigma)/2$.
In terms of scattering amplitude, the entanglement power vanishing conditions  are, (1) $\delta_1=\delta_0=0$ i.e.    $\mathcal{M}_1=\mathcal{M}_0=0$; (2) $\delta_1=\delta_0=\pi/2$ i.e.    $\mathcal{M}_1=\mathcal{M}_0=2i$; (3) $\delta_1=0,\delta_0=\pi/2$ i.e.    $\mathcal{M}_1=0,\mathcal{M}_0=2i$; (4) $\delta_1=\pi/2,\delta_0=0$ i.e.    $\mathcal{M}_1=2i,\mathcal{M}_0=0$. We note that in general,  the phase shifts, $\delta_1,\delta_0$, depend on the center of mass energy and the scattering angle, so these conditions hold for fixed kinematics. Out of the four entanglement suppression conditions, only $\mathcal{M}_1=\mathcal{M}_0=0$ is accessible in perturbative SMEFT, and the other three conditions require a resonance saturating the partial wave unitarity limit. Similar entanglement suppression conditions are realised in Table.~\ref{tab:suppression}. 
 For instance, in the backward scattering limit of  $H\tilde{H} \to H\tilde{H}$ channel, we obtain $\mathcal{M}_1=\mathcal{M}_0=0$ for which $S\propto I$. There are  three other entanglement suppression conditions for which $\mathcal{M}_1=\mathcal{M}_0$ but they are non-zero in general.

\section{Conclusions}\label{sec:conclusions}
We study the entanglement induced in $2\to2$ Higgs scattering in the unbroken phase of the electroweak symmetry, within the SMEFT framework. We regard the weak isospin of the SM Higgs doublet as a qubit, so the two outgoing Higgs fields form two isospin
qubits. Considering the momentum-isospin bipartition of the final state we obtain the momentum reduced density matrix, and we use the von Neumann entropy to measure entanglement between momentum and isospin. From the momentum reduced state, we compute the concurrence to quantify the entanglement between the two isospins. 

The $SU(2)_L$ symmetry of the Lagrangian  allows us to largely simplify the scattering matrix in terms of the
isospin singlet amplitude $\mathcal{M}_0$ and the triplet amplitude
$\mathcal{M}_1$. Therefore, every entanglement measure we compute is related to these two amplitudes together with the initial isospin configuration. Owing to $SU(2)_L$ symmetry, the scattering process does not change the initial entanglement when the initial state is an isospin singlet/triplet state. Non-trivial entanglement is induced by scattering when the initial state is an admixture of the isospin singlet and triplet states.

We analyse the variation of von Neumann entropy as a function of the total energy. It grows with energy due to the contribution from the dimension-6 and dimension-8  operators. The interference between different operators induces cancellation at a specific energy depending on the values of the Wilson coefficients.  
Assuming the leading contribution from the dimension-8 operators, we find relations among the Wilson coefficients for entanglement suppression in the forward or backward scattering limit. We revisit positivity bounds on dimension-8 operators and show their correlations with the conditions for entanglement suppression or generation. We also briefly comment on the entanglement power and derive  conditions for entanglement suppression, without restriction to specific initial states.

 It will be interesting to investigate whether the entanglement suppression conditions identified here have counterparts in the broken phase of the electroweak theory, where they would manifest as the interactions among physical Higgs and longitudinal components of weak gauge bosons.
 Finally, it would also be interesting to explore the impact of heavy mediators in various UV complete theories such as singlet scalar, triplet scalar, and massive graviton.
\section*{Acknowledgements}
This research was supported by the Chung-Ang University Research Grants in 2025. The work is supported in part by Basic Science Research Program through the National
Research Foundation of Korea (NRF) funded by the Ministry of Education, Science and
Technology (NRF-2022R1A2C2003567). RP acknowledges the valuable discussions with K. Kowalska and E.M. Sessolo at the National Centre for Nuclear Research, Warsaw, Poland. We appreciate the fruitful discussion with the organizers and participants during the CERN-CKC Theory workshop on Quantum Observables for Collider Physics 2026, where this work was further developed. 
\appendix 
\section{Feynman rules}\label{feynrule}
The Feynman rules for the vertices involving four Higgs scalars  are defined as  $i \lambda_{s_1 s_2 s_3 s_4} $, where $ p_i$ is the incoming momentum for scalar  $s_i$. The couplings are given by
\begin{equation}
\begin{aligned}
  i \lambda_{H^0 H^0 H^{0\star} H^{0\star}}&=i \lambda_{H^+ H^+ H^- H^-}\\&=
   -4 i ~\mathcal{C}_{H4}  - \frac{i ~\mathcal{C}_{H4D2} }{\Lambda ^2} \left( p_1.p_3 + p_1.p_4 +p_2.p_3 + p_2.p_4 \right) \\
   &-\frac{2i ~\mathcal{C}_{H \square }}{\Lambda ^2}  \left( p_1.p_1 +p_1.p_3 + p_1.p_4 + p_2.p_2 
  +  p_2.p_3+ p_2.p_4  + p_3.p_3 +  p_4.p_4 \right)
  \\ &   +\frac{2 i~\mathcal{C}_{1}}{\Lambda ^4} \left( p_1.p_4 ~p_2.p_3 + p_1.p_3 ~ p_2.p_4 \right) \\ &
  +\frac{4 i ~\mathcal{C}_{2} ~p_1.p_2 ~p_3.p_4}{\Lambda ^4}+\frac{2 i~\mathcal{C}_{3}}{\Lambda ^4} \left( p_1.p_4 ~p_2.p_3 + p_1.p_3 ~p_2.p_4 \right),
\end{aligned}
\end{equation}
and
\begin{equation}
\begin{aligned}
  i \lambda_{H^0 H^0 H^+ H^-}= &-2i~\mathcal{C}_{H4} -\frac{i~\mathcal{C}_{ H4D2}}{\Lambda ^2} (p_2.p_3 + p_1.p_4) \\
&-\frac{i~\mathcal{C}_{H\square }}{\Lambda ^2} \left(p_1.p_1+2 p_1.p_2+p_2.p_2+p_3.p_3+2 p_3.p_4+p_4.p_4\right) \\
&- \frac{2 i~\mathcal{C}_{1}}{\Lambda ^4} p_1.p_4 ~ p_2.p_3 -\frac{2 i~\mathcal{C}_{2}}{\Lambda ^4} p_1.p_3 ~p_2.p_4 -\frac{2 i~\mathcal{C}_{3}}{\Lambda ^4} p_1.p_2~ p_3.p_4 .
\end{aligned}
\end{equation}

In the general isospin basis for Higgs scalars, the Feynman rules for Higgs scattering, $H_\alpha H_\beta\to H_\gamma H_\delta$, are given by $i {\cal M}_{\alpha\beta,\gamma\delta}$, with
\begin{equation}
\begin{aligned}
{\cal M}_{\alpha\beta,\gamma\delta}=& -\frac{1}{2}C_{H4} \Big(\delta^{\gamma\alpha}\delta^{\delta\beta}+\delta^{\delta\alpha}\delta^{\gamma\beta} \Big) \\
&-\frac{1}{2}\frac{C_{H4 D2}}{\Lambda^2} \bigg[ \delta^{\gamma\alpha}\delta^{\delta\beta} (p_\gamma\cdot p_\beta +p_\delta\cdot p_\alpha)+\delta^{\delta\alpha}\delta^{\gamma\beta} (p_\delta\cdot p_\beta+p_\gamma\cdot p_\alpha)\bigg] \\
&-\frac{1}{2}\frac{C_{H\Box}}{\Lambda^2} \bigg[  \delta^{\gamma\alpha}\delta^{\delta\beta} (p_\delta\cdot p_\beta +p_\gamma\cdot p_\alpha)+\delta^{\delta\alpha}\delta^{\gamma\beta} (p_\gamma\cdot p_\beta+p_\delta\cdot p_\alpha)\bigg] \\
&-\frac{1}{4} \frac{C_{H\Box}}{\Lambda^2} \Big(\delta^{\gamma\alpha}\delta^{\delta\beta}+\delta^{\delta\alpha}\delta^{\gamma\beta}  \Big) (p^2_\alpha+p^2_\beta+p^2_\gamma+p^2_\delta) \\
&+\frac{2{\cal C}_1}{\Lambda^4} \bigg[  \delta^{\gamma\alpha}\delta^{\delta\beta}  (p_\gamma\cdot p_\beta)(p_\delta \cdot p_\alpha)+ \delta^{\delta\alpha}\delta^{\gamma\beta} (p_\gamma\cdot p_\alpha)(p_\delta\cdot p_\beta)\bigg] \\
&+\frac{2{\cal C}_2}{\Lambda^4} \Big(\delta^{\gamma\alpha}\delta^{\delta\beta}+\delta^{\delta\alpha}\delta^{\gamma\beta} \Big) (p_\gamma\cdot p_\delta)(p_\alpha\cdot p_\beta) \\
&+\frac{2{\cal C}_3}{\Lambda^4}  \bigg[  \delta^{\gamma\alpha}\delta^{\delta\beta}  (p_\gamma\cdot p_\alpha)(p_\delta\cdot p_\beta)+ \delta^{\delta\alpha}\delta^{\gamma\beta} (p_\gamma\cdot p_\beta)(p_\delta\cdot p_\alpha)\bigg]. 
\end{aligned}
\end{equation}
Here, $p_{\alpha}$ is the incoming momentum for scalar $H_\alpha$, and we need to multiply a symmetry factor 2 to the above rules for $\alpha=\beta$ or $\gamma=\delta$.
Then, the Bose symmetry becomes manifest in the above Feynman rules under $\alpha\leftrightarrow\beta$ and  $p_\alpha\leftrightarrow p_\beta$ in the initial states or $\gamma\leftrightarrow\delta$ and  $p_\gamma\leftrightarrow p_\delta$ in the final states. 
\section{Positivity bounds in the complex basis}\label{App:positivity}
In the complex basis for the SM Higgs doublet, we take the general superposition of scalars for the initial state as
\bea
|{\rm in}\rangle= |a\rangle \otimes |b\rangle
\eea
with
\bea
|a\rangle&=& \sum_{\alpha=1}^2 x_\alpha |H_\alpha\rangle + \sum_{{\bar\alpha}=1}^2 z_{\bar\alpha} |H^*_{\bar\alpha} \rangle, \\
|b\rangle &=& \sum_{\alpha=1}^2 y_\alpha |H_\alpha\rangle + \sum_{{\bar\alpha}=1}^2 w_{\bar\alpha} |H^*_{\bar\alpha} \rangle.
\eea
Then, we can impose positivity bounds for the elastic scattering, $ab\to ab$, by
\bea
{\overline M}(ab\to ab)\equiv \frac{1}{2} \frac{\partial^2}{\partial s^2} M(ab\to ab)(s,t=0)\geq 0, \label{positivity}
\eea
with
\bea
M(ab\to ab)&=&x_\alpha y_\beta x^*_\gamma y^*_\delta M_{\alpha\beta,\gamma\delta}(s,t,u) + z_{\bar\alpha} w_{\bar\beta} z^*_{\bar\gamma} w^*_{\bar\delta} M_{{\bar\alpha}{\bar\beta},{\bar\gamma}{\bar\delta}}(s,t,u)  \nonumber \\
&&+ x_\alpha w_{\bar \beta} x^*_{\gamma} w^*_{\bar\delta} M_{\alpha{\bar\beta},\gamma{\bar\delta}} (s,t,u) +z_{\bar\alpha} y_{ \beta} z^*_{\bar\gamma} y^*_{\delta} M_{{\bar\alpha}{\beta},{\bar\gamma}{\delta}}(s,t,u)   \nonumber \\
&&+z_{\bar\alpha} y_{ \beta} x^*_{\gamma} w^*_{\bar\delta} M_{{\bar\alpha}{\beta},\gamma{\bar\delta}}(s,t,u)  +x_{\alpha} w_{ \bar\beta} z^*_{\bar\gamma} y^*_{\delta} M_{{\alpha}{\bar\beta},{\bar\gamma}{\delta}}(s,t,u).
\eea
Here, $M_{\alpha\beta,\gamma\delta}(s,t,u),  M_{{\bar\alpha}{\bar\beta},{\bar\gamma}{\bar\delta}}(s,t,u)$, etc, correspond to the reduced scattering amplitudes for $H_\alpha H_\beta\to H_\gamma H_\delta$, $H^*_{\bar\alpha} H^*_{\bar\beta}\to H^*_{\bar\gamma} H^*_{\bar\delta}$, etc. Then, the crossing symmetry can be used to write down all the scattering amplitudes in terms of  $M_{\alpha\beta,\gamma\delta}(s,t,u)$, as follows,
\bea
M_{{\bar\alpha}{\bar\beta},{\bar\gamma}{\bar\delta}}(s,t,u) &=& M_{\gamma\delta,\alpha\beta}(s,t,u), \\
 M_{\alpha{\bar\beta},\gamma{\bar\delta}} (s,t,u) &=&M_{\alpha\delta,\beta\gamma}(s\to u,t\to s,u\to t), \\
 M_{{\bar\alpha}{\beta},{\bar\gamma}{\delta}}(s,t,u)  &=& M_{\beta\gamma,\alpha\delta}(s\to u,t\to s,u\to t), \\
 M_{{\bar\alpha}{\beta},\gamma{\bar\delta}}(s,t,u) &=& M_{\beta\delta,\alpha\gamma}(s\leftrightarrow t), \\
 M_{{\alpha}{\bar\beta},{\bar\gamma}{\delta}}(s,t,u) &=&  M_{\alpha\gamma,\beta\delta}(s\leftrightarrow t).
\eea
Here, we note that other amplitudes such as $M_{{\bar\alpha}\beta,\gamma \delta}, M_{\alpha\beta,{\bar\gamma}{\bar\delta}}$, etc, vanish identically, because of the hypercharge conservation.

For the dimension-8 operators with four Higgs fields, the reduced scattering amplitudes, $M_{\alpha\beta,\gamma\delta}(s,t,u)$, are given by
\bea
M_{11,11} &=&\frac{1}{2}\Big(\mathcal{C}_{1}+\mathcal{C}_{3}\Big)(u^2+t^2)+\mathcal{C}_{2}s^2 = M_{22,22},\\
M_{12,12}&=&\frac{1}{2}\Big(\mathcal{C}_{1} u^2+\mathcal{C}_{2} s^2+\mathcal{C}_{3} t^2\Big)  = M_{21,21}, \\
M_{12,21}&=&\frac{1}{2}\Big(\mathcal{C}_{1} t^2+\mathcal{C}_{2} s^2+\mathcal{C}_{3} u^2\Big)    = M_{21,12}.
\eea

Taking $ x_\alpha,  y_\alpha, z_{\bar\alpha}, w_{\bar\alpha}$ as being real, we express  ${\overline M}(ab\to ab)$ in eq.~(\ref{positivity}) as
\bea
{\overline M}(ab\to ab)=\mathcal{C}_{1}(X+Y)+\mathcal{C}_{2} (X+Y+Z)+ \mathcal{C}_{3}Y \geq 0, \label{superampl}
\eea
with
\bea
X &=& \frac{1}{2}(w_2 x_1-w_1 x_2)^2+ \frac{1}{2}(x_2 y_1- x_1 y_2)^2 \nonumber \\
&&+ \frac{1}{2} (w_2 z_1-w_1 z_2)^2+ \frac{1}{2}(y_2 z_1-y_1 z_2)^2,  \\
Y &=& \frac{1}{2}(w_1 x_1+w_2x_2+ y_1 z_1+y_2 z_2)^2 + \frac{1}{2}(x_1 y_1+ x_2 y_2+w_1 z_1+w_2 z_2)^2, \\
Z &=& \frac{1}{2}(w_1 x_1+w_2 x_2- y_1 z_1-y_2 z_2)^2+\frac{1}{2}(x_1 y_1+x_2 y_2-w_1 z_1-w_2 z_2)^2.
\eea
Therefore, as $X, Y, Z\geq 0$, imposing the positivity bounds \cite{Bi:2019phv,Remmen:2019cyz,Zhang:2021eeo,Yamashita:2020gtt,Kim:2023pwf,Kim:2023bbs} in eq.~(\ref{positivity}) with eq.~(\ref{superampl}) gives rise to
\bea
&&(1)\,\, \mathcal{C}_{1}+\mathcal{C}_{2}\geq 0, \\
&&(2)\,\,\mathcal{C}_{2} \geq 0, \\
&&(3)\,\, \mathcal{C}_{1}+\mathcal{C}_{2}+\mathcal{C}_{3}\geq 0.
\eea

For instance, the representative initial states giving rise to the positivity bounds are the following,
\bea
&&(1)\,\, |a\rangle=\frac{1}{\sqrt{2}}(|H_1\rangle+|H_2\rangle), \quad |b\rangle =\frac{1}{\sqrt{2}}(|H_1\rangle-|H_2\rangle), \label{App:positiv1} \\
&&(2)\,\,  |a\rangle=\frac{1}{\sqrt{2}}(|H_1\rangle+|H^*_1\rangle), \quad |b\rangle =\frac{1}{\sqrt{2}}(|H_1\rangle-|H^*_1\rangle), \label{App:positiv2}  \\
&&(3)\,\,  |a\rangle=\frac{1}{\sqrt{2}}(|H_1\rangle+|H^*_1\rangle)= |b\rangle.  
\label{App:positiv3}
\eea
Thus, it is sufficient to take the qubit basis for the initial state for each positivity bound: $\{|H_1\rangle,|H_2\rangle\}$ for $(1)$ and $\{|H_1\rangle,|H^*_1\rangle\}$ for $(2)$. We note that there exist alternative qubit basis leading to the same positivity bounds, but it is sufficient to focus on the above representative qubit basis to make connection between the positivity bounds and the entanglement.

\medskip
\bibliographystyle{apsrev4-1}
\bibliography{ref}
\end{document}